\documentclass[a4paper,twocolumn,10pt,accepted=2021-03-30]{quantumarticle}
\pdfoutput=1

\usepackage{amssymb}
\usepackage{mathtools}
\usepackage{graphicx}
\usepackage{bbm}
\usepackage{color}
\usepackage{comment}
\usepackage{gensymb}
\usepackage{cite}
\usepackage[numbers,sort&compress]{natbib}
\definecolor{blue}{rgb}{0,0,1}
\definecolor{grey}{rgb}{0.6,0.6,0.6}

\newcommand{\bra}[1]{\langle #1 |}
\newcommand{\ket}[1]{| #1 \rangle}

\begin{document}

\title{Qubit-efficient encoding schemes for binary optimisation problems}

\author{Benjamin Tan}
\email[]{b.tan@.nus.edu}
\affiliation{Centre for Quantum Technologies, National University of Singapore, 3 Science Drive 2, Singapore 117543}
\author{Marc-Antoine Lemonde}
\email[]{cqtml@nus.edu.sg}
\affiliation{Centre for Quantum Technologies, National University of Singapore, 3 Science Drive 2, Singapore 117543}
\author{Supanut Thanasilp}
\affiliation{Centre for Quantum Technologies, National University of Singapore, 3 Science Drive 2, Singapore 117543}
\author{Jirawat Tangpanitanon}
\affiliation{Centre for Quantum Technologies, National University of Singapore, 3 Science Drive 2, Singapore 117543}
\author{Dimitris G. Angelakis}
\email[]{dimitris.angelakis@qubit.org}
\affiliation{Centre for Quantum Technologies, National University of Singapore, 3 Science Drive 2, Singapore 117543}
\affiliation{School of Electrical and Computer Engineering, Technical University of Crete, Chania, Greece 73100}


\begin{abstract}
We propose and analyze a set of variational quantum algorithms for solving quadratic unconstrained binary optimization problems where a problem consisting of $n_c$ classical variables can be implemented on $\mathcal O(\log n_c)$ number of qubits. 
The underlying encoding scheme allows for a systematic increase in correlations among the classical variables captured by a variational quantum state by progressively increasing the number of qubits involved. 
We first examine the simplest limit where all correlations are neglected, i.e.~when the quantum state can only describe statistically independent classical variables.
We apply this minimal encoding to find approximate solutions of a general problem instance comprised of 64 classical variables using $7$ qubits. 
Next, we show how two-body correlations between the classical variables can be incorporated in the variational quantum state and how it can improve the quality of the approximate solutions. We give an example by solving a $42$-variable Max-Cut problem using only $8$ qubits where we exploit the specific topology of the problem. 
We analyze whether these cases can be optimized efficiently given the limited resources available in state-of-the-art quantum platforms.
Lastly, we present the general framework for extending the expressibility of the probability distribution to any multi-body correlations.
Our encoding scheme allows current generations of quantum hardware to explore the boundaries of classical intractability involving real world problem sizes, and can be implemented on currently available quantum hardware across various platforms, requiring only few dozens of qubits.
\end{abstract}

\maketitle

\section{Introduction}
In recent years, important experimental breakthroughs have propelled quantum computing as one of the most thriving fields of research~\cite{Saffman2016, Wendin2017, Bruzewicz2019, Google2019QS}, with the long-term goal of building universal quantum computers capable of running algorithms with provable quantum speed-up~\cite{Shor1997, Grover1997}. 
As the first generations of quantum hardware, referred to as noisy intermediate-scale quantum (NISQ) devices~\cite{Preskill2018NISQ}, do not yet fulfill the technical requirements to implement error-corrected universal quantum computing, increasing efforts are dedicated to design near-term algorithms capable of performing computational tasks with imperfect and limited quantum resources~\cite{Bauer2020, McArdle2020}. 
Amongst the most promising paradigms are the variational quantum algorithms (VQA)~\cite{Farhi2014QAOA1, Peruzzo2014, McClean2016, Kandala2017, Moll2018}.
In these algorithms, a parameterized quantum circuit is optimized using classical computing resources to generate a quantum state that represents an accurate approximate solution of the problem at hand. 
While a formal proof of any quantum advantages these algorithms might bring has yet to be found~\cite{Ronnow2014}, applications of NISQ devices to real-world problems are already being explored in chemistry~\cite{McArdle_2020} and unconstrained binary optimisation (QUBO) problems~\cite{Garey1979}.

The QUBO model is an NP-hard combinatorial problem that consists of minimizing a cost function of the form $C_{x} = \vec x^\intercal \mathcal A \vec x$, where $\vec x\in\{0,1\}^{n_c}$ is a vector of $n_c$ classical binary variables and $\mathcal A$ is a real and symmetric matrix.
VQAs such as the quantum approximate optimization algorithm (QAOA)~\cite{Farhi2014QAOA1, Farhi2014QAOA2, Farhi2019, Otterbach2017, Qiang2018, Pagano2019, Bengtsson2019, Abrams2019, Willsch2020, Arute2020} and hardware efficient~\cite{Rattew2019, Braine2019} approaches have been applied to find approximate solutions to QUBO problems. QAOA in particular is able to ensure a lower bound on the quality of its solutions for sparse instances of QUBO problems using shallow circuits. This quality is then able to monotonically converge towards the exact solution for infinitely deep circuits recovering quantum adiabatic computing~\cite{Farhi2014QAOA1, Zhou_2020}. 
Recent experiments have however highlighted the challenges in implementing the QAOA on problem graphs that differ from the native connectivity of the quantum hardware for increasing system sizes~\cite{Arute2020}.
Hardware efficient approaches, on the other hand, are motivated by the simplicity of their implementations but do not guarantee a lower bound on the quality of the solutions.
This is usually accomplished using series of gates native to the quantum platform and is unconstrained by the topology of the QUBO problem.
However, depending on the implementation, these hardware efficient approaches can be plagued by exponentially large barren plateaus in their optimization landscapes as the number of qubits increases~\cite{McClean2018}.
In addition to increasing algorithmic difficulties, the engineering overhead of scaling up the quantum hardware also currently limits the size of computational tasks to toy models.
Previously proposed schemes to implement quantum algorithms to solve optimization problems have used a number of classical variables equal to the number of qubits available and were therefore limited to problem sizes involving only a few tens of them~\cite{Otterbach2017, Qiang2018, Pagano2019, Bengtsson2019, Abrams2019, Willsch2020, Arute2020, Rattew2019, Braine2019}. This is not representative of real-world optimization problems, where the number of classical variables $n_c$ involved can be on the order of $10^4$. 

In this work, we tackle this problem by proposing an encoding scheme for QUBO models with $n_c$ variables that can be implemented on $\mathcal O(\log n_c)$ number of qubits.
We devise a strategy using $n_a$ ancilla qubits and $n_r$ register qubits to divide the QUBO problem into $2^{n_r}$ subsystems of $n_a$ classical variables, requiring a total of $n_q = n_a + n_r$ qubits.
This approach allows for a simultaneous search through each subsystem by exploiting the intrinsic parallelism offered by quantum devices.
In this context, the resulting variational quantum state that encodes a probability distribution over all classical solutions is capable of capturing any $n_a$-body correlations between a number of QUBO variables that scale exponentially with $n_r$.
This heuristic approach allows for the systematic increase in the correlations that can be captured in the probability distribution by progressively increasing the number of qubits. 
As an example, in the limit where each subsystem is composed of only a single classical variable, i.e.~all correlations between classical variables are neglected, $n_r = \log_2(n_c)$ and optimization problems of size $n_c\sim10^4$ could be tackled on quantum hardware with no more than 15 qubits.
We emphasize that this limiting case of $n_a=1$ can be efficiently classically simulated and therefore should not provide any quantum speed-up.
At the other end of the spectrum, the most resource intensive limit of our encoding scheme is reached when $n_q = n_a = n_c$ and recovers the traditional approaches that are classically intractable and possibly offer quantum advantages.
This encoding scheme provides a systematic way to traverse between these limits, thus allowing one to balance between capturing selected amounts of correlations whilst respecting the hardware capabilities of modern day devices.
With the expected capabilities of upcoming NISQ devices, this scheme paves the way to explore the boundaries of classical intractability for real-world problem sizes.

In what follows, we introduce the general idea of our systematic encoding scheme and numerically demonstrate how the limit of $n_a=1$ is able to solve QUBO problems while significantly reduce the number of qubits required.
From there, we make the first step towards more expressive encoding by considering protocols to capture different subsets of two-body correlations and explore whether they can be optimized efficiently.
We demonstrate numerically how a selective encoding scheme can be applied to the Max-Cut problem and show that exploiting the topology of a specific problem to select an efficient subset of correlations leads to better solutions.
All protocols proposed in this manuscript are in line with the limitations of the current state-of-the-art quantum platforms.


\section{QUBO model and the complete encoding scheme}

The QUBO model is an NP-hard combinatorial problem that consists of minimizing a cost function of the form $C_{x} = \vec x^\intercal \mathcal A \vec x$, where $\vec x\in\{0,1\}^{n_c}$ is a vector of $n_c$ classical binary variables and $\mathcal A$ is a real and symmetric matrix.
This model is of particular interest due to its relationship with other optimization problems such as the Max-Cut, portfolio optimization and facility allocation problems \cite{Karp1972, Markowitz1952, Koopmans1957}. 
Existing metaheuristic approaches such as the TABU search, genetic algorithms, and simulated annealing are capable of finding suitable solutions to problems consisting of $n_c \sim 10^4$ classical variables~\cite{Fouskakis2002, kochenberger2014}.

In recent implementations of VQA applied to solving QUBO problems, each binary variable in $\vec x$ is represented by a single qubit, i.e.~$n_q = n_c$; a mapping which we will refer to as the {\it complete encoding}. The resulting quantum state is parameterized by a set of angles $\vec \theta$ with the general form
\begin{equation}
	\ket{\psi_{\textrm{cp}}(\vec \theta)} = \hat U_{\textrm{cp}}(\vec \theta)\ket{\psi_0} = \sum_{\vec x\in\{0,1\}^{n_c}} \alpha_{\vec x}(\vec \theta) \ket{\vec x}, \label{Eq:FullEncoding}
\end{equation} 
where $\hat U_{\textrm{cp}}(\vec \theta)$ is the unitary evolution implemented on the quantum platform, $\{ \ket{\vec x} = \otimes_i^{n_q}\ket{x_i}\}$ with $x_i\in\{0,1\}$ is the complete computational basis spawn by the $n_q$ qubits and $\ket{\psi_0}$ is a given input state.
By associating a classical solution $\vec x$ with a basis state $\ket{\vec x}$, the state $\ket{\psi_{\textrm{cp}}(\vec \theta)}$ is able to encode all possible classical solutions in a linear superposition. This unique property of quantum mechanics opens the possibility for multiple classical solutions to be tested simultaneously and this intrinsic parallelism is a strong motivator in developing quantum algorithms for classical problems.

In the case where $\hat U_{\textrm{cp}}(\vec \theta)$ is a universal quantum circuit, all $\alpha_{\vec x}$ in Eq.~\eqref{Eq:FullEncoding} can in principle be independent (up to the normalization condition). Consequently, this quantum state is able to capture all possible correlations between the classical variables and exhibits expressive power that is beyond classical computation~\cite{Du2018, Killoran2019, Coyle2019}.
The goal from here would be to efficiently navigate the exponentially large Hilbert space and reach the basis state(s) which represent the exact or approximate solution(s) to the QUBO problem.


\begin{figure}    
\center
\includegraphics[width=0.85\columnwidth]{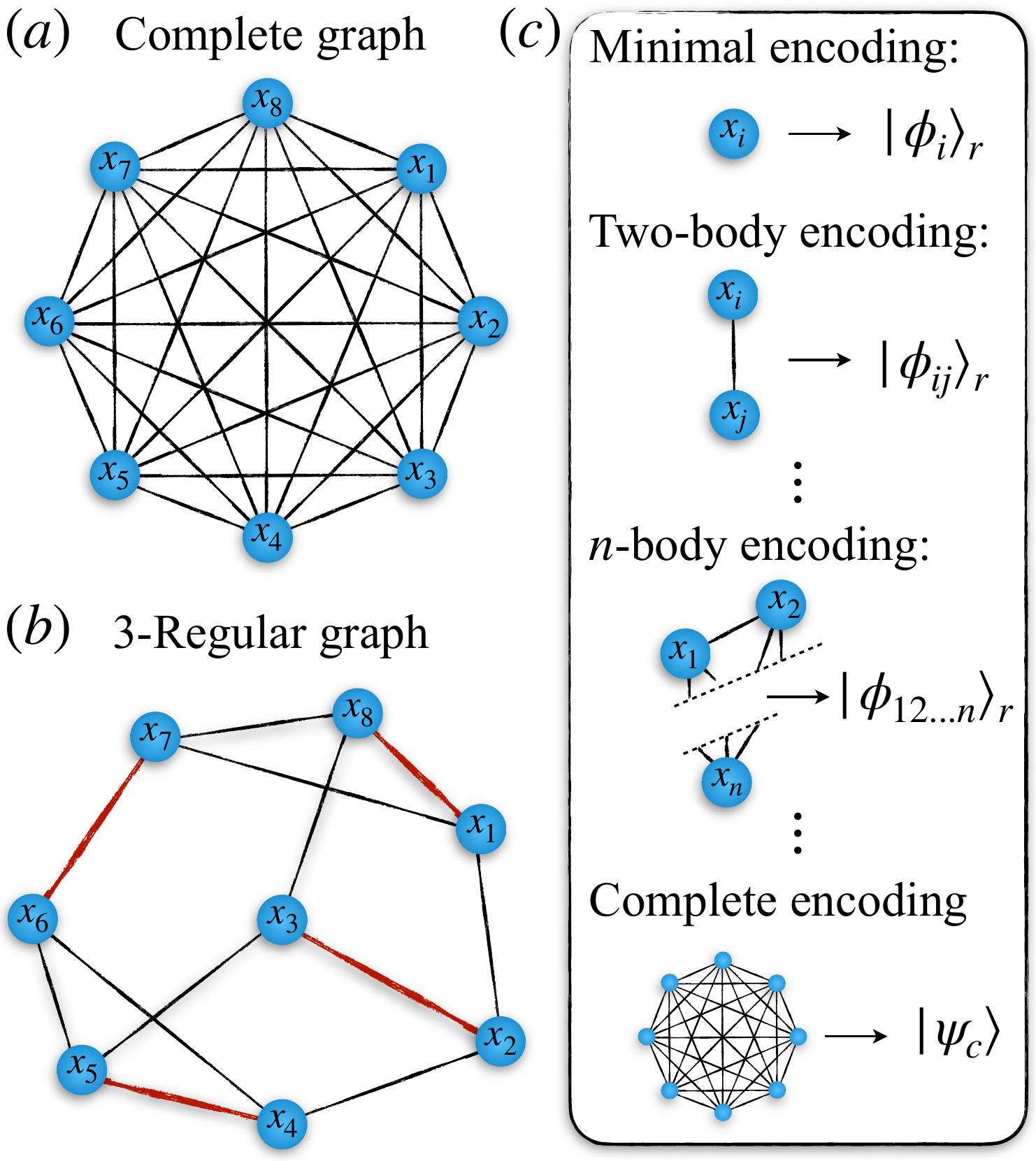} 
\caption{Schematic representation of the encoding schemes. (a) Complete graph representing a dense QUBO matrix $\mathcal A$. (b) A 3-Regular graph Max-Cut problem. The set of red edges is an example of perfect matching where each vertex is connected to only a single edge. (c) Encoding schemes where $n_a$-body correlation are encoded with $n_a$ increasing from top to bottom. In the minimal encoding, each of the $2^{n_r}$ basis states $\ket{\phi_i}$ is used to represent a single classical variable $x_i$ (vertex). In the $n$-body (two-body) encoding scheme, groups of $n$ (two) classical variables are formed and each basis state represents a unique encoded group. In the complete encoding, each basis state represents an entire graph.} \label{Fig:Schema}
\end{figure}


In this complete encoding scheme, the QUBO model can be mapped onto an Ising Hamiltonian 
\begin{equation} \label{Eq:HIsing}
\hat H_{\rm Ising} = \frac{1}{4}\sum_{i,j}^{n_c} \mathcal A_{ij}(1-\hat\sigma_z^{(i)})(1-\hat\sigma_z^{(j)}),
\end{equation}
where $\hat\sigma_z^{(i)}$ is the $z$ Pauli matrix acting on qubit $i$ and $\mathcal A_{ij}$ are the elements of the matrix $\mathcal A$.
The ground state of $\hat H_{\rm Ising}$ is a basis state $\ket{\vec x}$ that corresponds to an exact solution $\vec x$ of the QUBO problem defined by $\mathcal A$.
For general instances, $\hat H_{\rm Ising}$ represents a system of interacting spins where all two-body interactions may be present.

A variational algorithm can then be implemented to find a suitable solution by using the ansatz $\hat U_{\textrm{cp}}(\vec \theta)$ to produce trial states and finding the set of parameters $\vec\theta$ to minimize the cost function
\begin{equation} \label{Eq:CF_FullEnc}
	C_{\textrm{cp}}(\vec \theta) = \bra{\psi_{\textrm{cp}}(\vec \theta)}\hat H_{\rm Ising}\ket{\psi_{\textrm{cp}}(\vec \theta)}.
\end{equation} 
Here, Eq.~\eqref{Eq:CF_FullEnc} is a linear function of expectation values with a number of terms polynomial in the number of qubits. 

Existing variational ansatzes for optimization problems can be divided in two distinct groups --- approaches which require the Hamiltonian $\hat H_{\rm Ising}$ to be implemented on the quantum hardware and those which utilize only native gates unconstrained by the specific problem. 
Approaches such as the QAOA, as implemented in Refs.~\cite{Otterbach2017, Qiang2018, Pagano2019, Bengtsson2019, Abrams2019, Willsch2020, Arute2020}, fall into the first category and benefit from being able to exploit some extent of adiabatic computing to search the Hilbert space~\cite{AQCReview2018}. 
In principle, the produced variational state is guaranteed to converge towards the exact solution for infinitely long quantum evolution $\hat U_{\textrm{cp}}(\vec\theta)$. These approaches however, can be difficult to implement for generic QUBO problems. 
Approaches that fall into the second category have been implemented in Refs.~\cite{Rattew2019, Braine2019} and are designed to circumvent the technical challenges of implementing $\hat H_{\rm Ising}$. However, these approaches do not guarantee the existence of an efficient path to the optimal solution and can become exponentially hard to optimize as the system size increases.
While the ansatz proposed in this work belongs to the latter category, there should be no fundamental restrictions on devising circuit structures tailored toward a specific QUBO problem within the proposed encoding schemes.


\section{Minimal encoding}

While complete encoding schemes allow for all many-body correlations to be captured between classical variables, the number of qubits required limits their application to small system sizes with unfavorable scaling up perspectives. 
In what follows, we propose an encoding scheme which sacrifices this ability to capture correlations but allows for problem sizes to be scaled exponentially with the number of qubits.
We refer to this mapping as the {\it minimal encoding}.

\subsection{Expressibility of the minimal encoding}
The minimal encoding scheme considered here requires one ancilla $n_a = 1$ and $n_r = \log_2 n_c$ register qubits for a total number of $n_q = \log_2 n_c + 1$ qubits.  
The parametrized quantum state can be expressed as
\begin{equation}
	\ket{\psi_1({\vec \theta})} = \sum_{i=1}^{n_c} \beta_i(\vec\theta) [a_i(\vec\theta) \ket{0}_a + b_i(\vec\theta) \ket1_a]\otimes\ket{\phi_i}_r, \label{Eq:MinEncoding}
\end{equation}
where the states $\{\ket{\phi_i}_r\}$ ($\{ \ket0_a , \ket1_a \}$) are computational basis states of the register (ancilla) qubits. The premise is to define a one-to-one correspondence between each of the $n_c$ classical variables $x_i$ in $\vec x$ and a unique basis state $\ket{\phi_i}_r$, as depicted in Fig.~\ref{Fig:Schema} (c). 
The probability of the $i^{\rm th}$ classical variable to have the value $1$ or $0$ is given by Pr$(x_i = 1) = |b_i|^2$ and Pr$(x_i = 0) = 1- |b_i|^2= |a_i|^2$ respectively.
The coefficients $\beta_i(\vec\theta)$ capture the likelihood of measuring each register state $\ket{\phi_i}$ and thus the corresponding state of the ancilla qubit.
As an example, encoding the probability distribution over all solutions $\vec x$ of dimensions $n_c=4$ requires $n_r=2$. One can then define the mapping as $\ket{\phi_1}_r \equiv \ket{00}_r$, $\ket{\phi_2}_r = \ket{01}_r$, $\ket{\phi_3}_r \equiv \ket{10}_r$ and $\ket{\phi_4}_r = \ket{11}_r$.
In doing so, the quantum state representing the unit probability of sampling $\vec x = (1,0,0,1)$ would read $\ket{\psi_1} = (\ket1_a\ket{00}_r + \ket0_a\ket{01}_r + \ket0_a\ket{10}_r + \ket1_a\ket{11}_r)/2$.
A similar encoding strategy has been utilized in the context of image compression~\cite{Latorre2005}.

The limitation of this compact mapping is its ability to only encode distribution functions of statistically independent classical variables, i.e.~where the probability of obtaining a particular classical solution $\vec{x}$ from the state is given by Pr$(\vec{x}) = \prod_{i=1}^{n_c}{\rm Pr}(x_i)$. 
This comes as no surprise as the quantum state uses only $n_c$ coefficients to encode a probability distribution over $2^{n_c}$ solutions. 
As a consequence, it is always possible to efficiently capture the resulting distribution functions using classical approaches, as we will discuss below in more detail. Despite these limitations, we examine this limiting case closely as it captures the core elements of the general encoding strategy.


	
\begin{figure}
\center
\includegraphics[width=0.95\columnwidth]{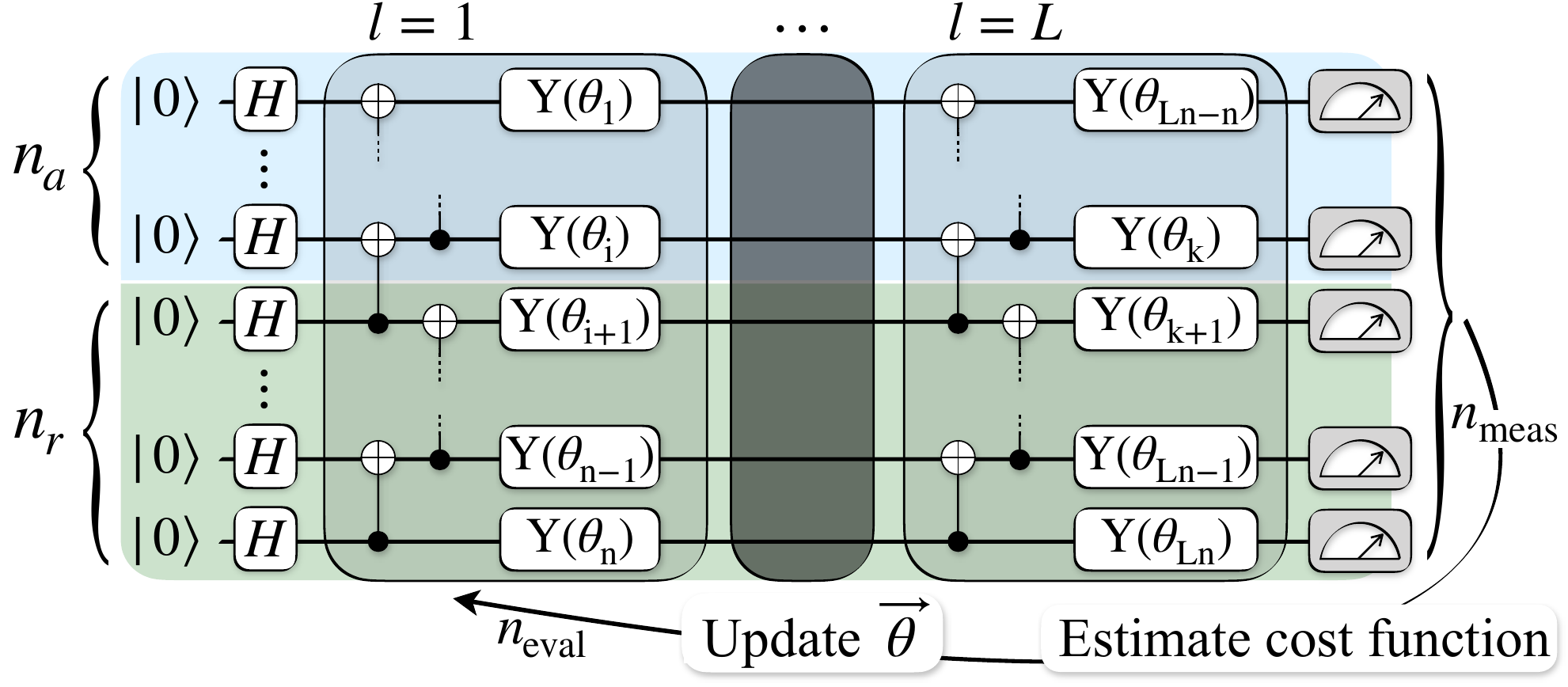}
\caption{Hardware-efficient variational ansatz. The initial quantum state $\ket{\psi_0} = \frac{1}{\sqrt{2^{n_q}}}\otimes_{i=1}^{n_q} (\ket0 + \ket1) $ is produced by the first layer of Hadamard gates. Each subsequent layer $1\leq l\leq L$ is composed of a series of CNOT gates and single rotations $R_y(\theta_{i})$ (denoted $\rm Y$) where the $L\times n_q$ variational parameters are grouped in $\vec\theta$ ($n=n_q$ for readability). 
A single evaluation of the cost function requires $n_{\rm meas}$ measurements in the computational basis and one optimization process requires $n_{\rm eval}$ of these evaluations.} 
\label{Fig:Circuit}
\end{figure}


\subsection{Cost function to minimize}

As with standard VQAs, we defined a cost function to be minimized using a set of parameters $\vec\theta$.
Given that $\ket{\psi_1(\vec\theta)}$ represents a distribution function over statistically independent classical variables, it adopts the form
\begin{align}
	C_1(\vec\theta) 
	& = \sum_{i,j=1}^{n_c} \mathcal A_{ij}  \frac{\langle \hat P_i^{1}\rangle_{\vec\theta}\langle\hat P_j^{1}\rangle_{\vec\theta}}{\langle \hat P_i\rangle_{\vec\theta}\langle\hat P_j\rangle_{\vec\theta}}(1-\delta_{ij})  +\sum_{i=1}^{n_c} \mathcal  A_{ii} \frac{\langle \hat P_i^{1}\rangle_{\vec\theta}}{\langle \hat P_i\rangle_{\vec\theta}},  
	\label{Eq:C1}
\end{align}
where $\hat P_i = \ket{\phi_i}\bra{\phi_i}_r$ ($\hat P^1_i = \ket{1}\bra{1}_a \otimes \hat P_i$) are the projectors over the register basis state $\ket{\phi_i}_r$ independent of the ancilla state (with the ancilla being in $\ket{1}_a$). The expectation value can be expressed as $\langle \hat P_i \rangle_{\vec\theta} = \bra{\psi_1(\vec\theta)}\hat P_i\ket{\psi_1(\vec\theta)}$, giving $b_i(\vec \theta) = \langle \hat P_i^{1}\rangle_{\vec\theta}/\langle \hat P_i\rangle_{\vec\theta}$.

The highly entangled quantum state that minimizes Eq.~\eqref{Eq:C1} adopts the form $\ket{\psi} = \sum_i \beta_i \ket{\sigma_i}_a\otimes\ket{\phi_i}$ with $\sigma_i = \{0,1\}$ and corresponds unambiguously to the exact solution $\vec x = [\sigma_1, \dots, \sigma_{n_c}]$ that minimizes the QUBO problem defined by the matrix $\mathcal A$.
This point is crucial as it ensures that finding the global minimum of Eq.~\eqref{Eq:C1} leads to the exact classical solution that minimizes the QUBO problem.
Another important aspect of $C_1(\vec \theta)$ is that it only depends on the set of norms $\{|b_i|^2\}$. As a consequence, partial tomography performed by a series of measurements solely in the computational basis is sufficient for its estimation.
Finally, the cost function $C_1(\vec\theta)$ in Eq.~\eqref{Eq:C1} cannot be reduced to a linear function of expectation values and therefore the QUBO model in the minimal encoding scheme cannot be described with a suitable Hamiltonian.
A detailed derivation of Eq.~\eqref{Eq:C1} is presented in appendix \ref{App:CostFunctions}.

	
\begin{figure}
\center
\includegraphics[width=0.9\columnwidth]{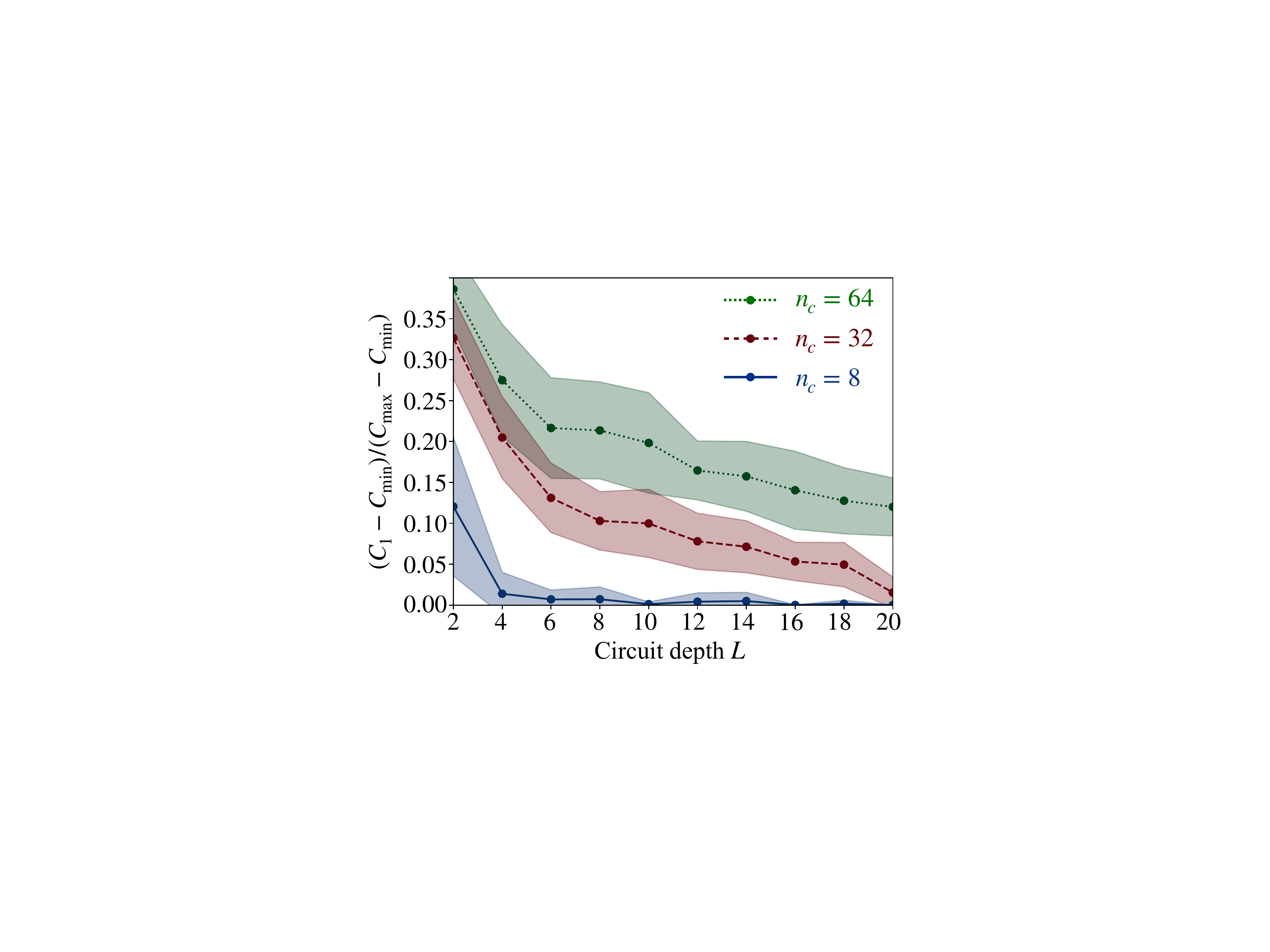}
\caption{
Optimized cost function value in the minimal encoding scheme as a function of circuit depth $L$ for randomly generated problems of different sizes $n_c$ with $n_{\rm meas} \rightarrow \infty$. The lines represent the mean final value of $20$ runs from $20$ different starting points $\vec \theta_{\rm ini}$. The shaded regions show one standard deviation from the mean. The optimization processes have been terminated at $n_{\rm eval} = 5000$. The minimum and maximum values of the cost function, $C_{\rm min}$ and $C_{\rm max}$ respectively, were found using the classical optimization software Gurobi~\cite{Gurobi} and were used to normalize the cost function values from 0 to 1.} 
\label{Fig:CF_vs_L_Na1}
\end{figure}

	 
\begin{figure*}
\center
\includegraphics[width=0.9\textwidth]{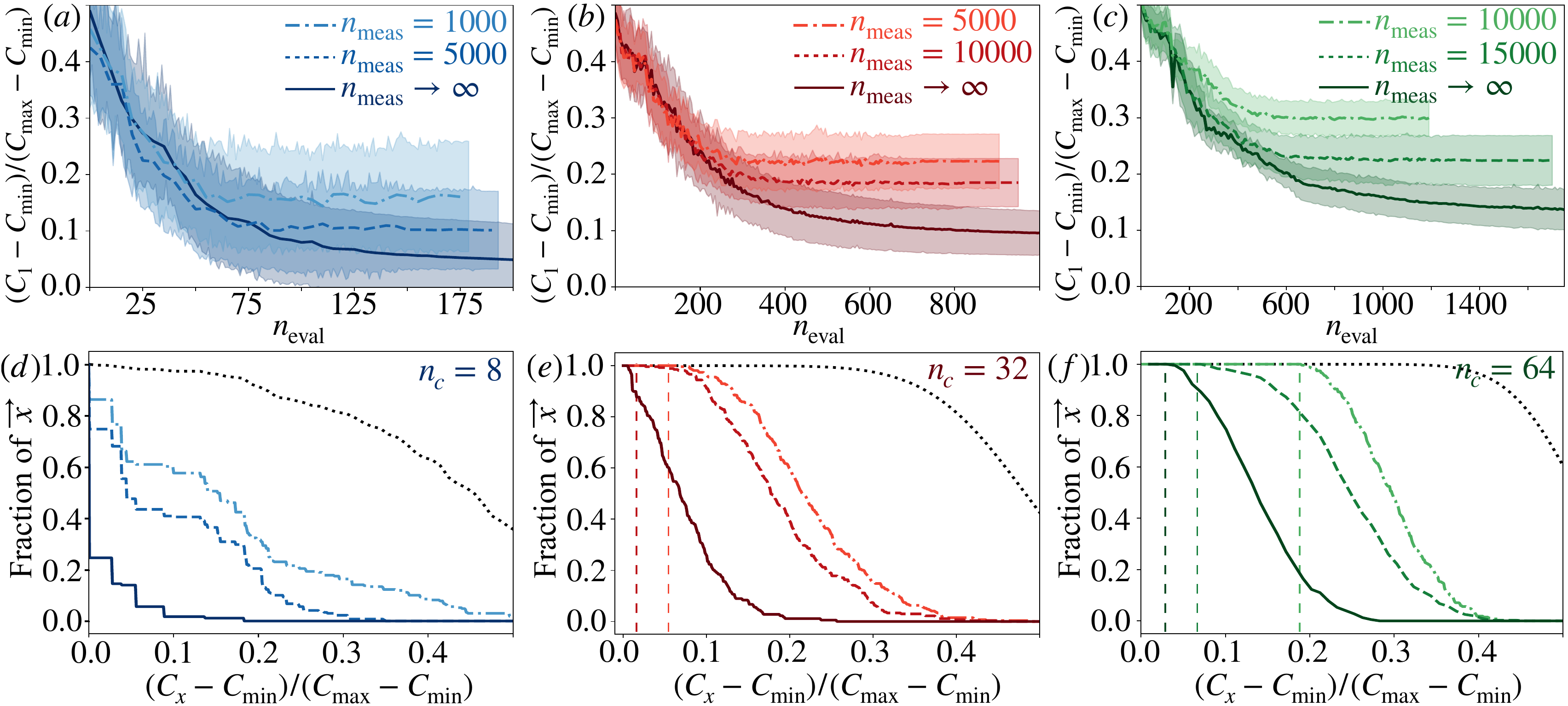}
\caption{
(a)--(c): Evolution of the (renormalized) cost function $\tilde C_1 = (C_1 - C_{\rm min})/ (C_{\rm max} - C_{\rm min})$ during the optimization process in the minimal encoding scheme for different sizes of $\mathcal A$. Shaded regions represent one standard deviation from the average over $30$ different starting $\vec\theta_{\rm ini}$. The final value of the cost function decreases as more $n_{\rm meas}$ were used in evaluating $\tilde C_1$. (a) $n_q=4$ qubits with circuit depth of $L=4$ and matrix size of $n_c = 8$. (b) $n_q=6$ qubits, circuit depth of $L=12$, and matrix size of $n_c = 32$. (c) $n_q=7$ qubits, circuit depth of $L=18$, and matrix size of $n_c = 64$. 
(d)--(f): Cumulative distribution of solutions drawn from the final quantum state $\ket{\psi_1(\vec \theta_{\rm opt})}$ using the parameters outlined in (a)--(c). $10$ classical solutions were drawn from each starting $\vec\theta_{\rm ini}$ after optimization.
The y-axis represents the fraction of solutions with value above the corresponding $\tilde C_x$ (x-axis). Vertical lines show the minimum $\tilde C_x$ found from the solutions. The black dotted line shows the distribution of (d) all $2^{n_c} = 256$ solutions and (e)--(f) $4\times 10^8$ randomly generated solutions.} 
\label{Fig:Training_Na1}
\end{figure*}


\subsection{Variational protocol to solve randomly generated QUBO models} \label{Sec:Training1q}

The quantum state $\ket{\psi_1(\vec \theta)} = \hat U_1(\vec\theta)\ket{\psi_0}$ is produced by a parameterized unitary evolution $\hat U_1(\vec\theta)$ applied to an initial product state $\ket{\psi_0} \sim (\ket0_a + \ket1_a) \otimes \sum_{i=1}^{n_c} \ket{\phi_i}_r$. 
We consider a hardware efficient circuit as our ansatz $\hat U_1(\vec\theta)$ in the form depicted in Fig.~\ref{Fig:Circuit}. 
This circuit starts with a layer of Hadamard gates applied to all the qubits initially in $\ket{00\dots00}$ to produce $\ket{\psi_0}$.
It then follows with an alternating sequence of nearest-neighbor CNOT gates and single qubit $R_y(\theta_i)$ rotations. Each successive application of CNOT gates and $R_y(\theta_i)$ rotations make up a single layer.
This choice of ansatz represents the simplest case where qubits are arranged in a linear topology with nearest-neighbor couplings. It also produces states with only real-valued coefficients which efficiently restricts the Hilbert space since the cost function in Eq.~\eqref{Eq:C1} does not depend on any phases. 

The optimization procedure is standard and first consists of randomly choosing a starting point for the variational parameters $\vec\theta_{\rm ini}$ from a uniform distribution and measuring the output quantum state $\ket{\psi(\vec\theta_{\rm ini})}$ in the computational basis. This quantum evolution is repeated $n_{\rm meas}$ times to estimate $C_1(\vec\theta_{\rm ini})$. The results are fed to a classical optimizer which updates the parameters $\theta_{\rm old} \rightarrow \theta_{\rm new}$. The parameters are updated $n_{\rm eval}$ times until convergence or if a set of termination criteria is met. The resulting parameters are denoted $\vec\theta_{\rm opt}$. From the final quantum state $\ket{\psi(\vec \theta_{\rm opt})}$, a set of solutions $\{\vec x\}$ with value $C_{x} = \vec x^\intercal A \vec x$ are obtained by sampling each variable independently following Pr$(x_i = 1) = |b_i|^2$ [cf.~Eq.~\eqref{Eq:MinEncoding}].

In Fig.~\ref{Fig:CF_vs_L_Na1}, we show the average optimized cost function as a function of circuit depth for $3$ QUBO instances of different sizes, $n_c = 8, 32$ and $64$, using $n_q = 4, 6$ and $7$ qubits respectively\footnote{We note that the expressive power of $\ket{\psi_1(\vec \theta)}$ can be fully captured within the complete encoding scheme by using only a single layer of $R_y(\theta_i)$ rotations applied to each qubit. Studying the minimal encoding scheme is therefore akin to examining the amount of resources required to map the simplest quantum circuit of $n_c$ qubits onto an exponentially narrower circuit.}. 
In each instance, the elements of $\mathcal A$ were randomly drawn from a uniform distribution ranging from -1 to 1.
COBYLA was chosen as the classical minimizer to update the variational parameters as it was found to give the best results for the least number of cost function evaluations~\cite{Powell2007}. The effects of a noisy circuit is shown in Appendix \ref{App:Noise} where we compare the performance of a noise-free optimization for a $n_c=32$ matrix to one with a simplified noise model applied.
In Fig.~\ref{Fig:Training_Na1} (a)--(c), we compare the infinite-measurement limit to simulated values obtained using $n_{\rm meas} \sim 1-15 \times 10^3$, at specific circuit depths for each of the different problem sizes.
Our findings show that increasing the number of measurements reduces the likelihood of the optimizer terminating in a local minima caused by fluctuations in the cost function.
It also allows for finer tuning of the optimal parameters due to the increased precision when estimating $C_1(\vec\theta)$, resulting in an increase in $n_{\rm eval}$. 

From each of the optimized states of Fig.~\ref{Fig:Training_Na1} (a)--(c), $10$ classical solutions $\vec x$ were drawn and distributed according to their normalized cost function value $\tilde C_x = (C_x - C_{\rm min})/ (C_{\rm max} - C_{\rm min})$; their normalized cumulative sum is shown in Fig.~\ref{Fig:Training_Na1} (d)--(f).
The resulting histogram $y(\tilde C_x)$ (y-axis) represents the fraction of the solutions drawn that have a cost function greater than $\tilde C_x$ (x-axis).
As an example, a value $y(\tilde C_x = 0.2) = 0.3$ means that $30\%$ of solutions drawn have a cost function value of $\tilde C_x > 0.2$.
The better the solutions $\{\vec x\}$ obtained, the sharper the histogram will peak at $\tilde C_x = 0$. 
The different coloured lines stand for different number of measurements $n_{\rm meas}$ and are compared to randomly drawn solutions as represented by the dotted black curves. 
The results show that the minimal encoding scheme was able to produce a significant portion of its solutions to be within $20\%$ of the optimal cost function value for $n_c = 8, 32$ and a majority of solutions produced for the $n_c = 64$ case were found to be within $30\%$ of the optimal cost function value.
The numerical results also suggest that an increase in resources such as $n_{\rm meas}$, $n_{\rm eval}$ and depth $L$ are required to maintain comparable accuracy as the problem sizes increase.

In Appendix \ref{App:Noise}, we investigate the robustness of the minimal encoding to experimental imperfections such as finite gate fidelities, coupling to the environment and readout errors. We further compare its performance to the more standard QAOA protocol in Appendix \ref{App:QAOA}. Following the recent state-of-the-art QAOA experiments for a fully connected problem in Ref.~\cite{Arute2020}, implementing the QAOA for $n_c = 8$ variables would require $8$ qubits and $612$ gates for $p=3$. In comparison, the minimal encoding with a hardware efficient ansatz only requires $4$ qubits and $42$ gates for $L=6$ and was able to achieve an improvement in performance over the QAOA. While current noise levels encountered in state-of-the-art experiments affect all quantum optimization algorithms proposed so far, we show that compared to the QAOA, our efficient encoding is a step in the right direction by drastically reducing the resources required to solve larger-scale problems.

\subsection{Classical Simulatability} \label{Sec:classicalsim}




As previously mentioned, the exponential decrease in the number of qubits offered by the minimal encoding also limits its advantage over classical methods. The probability distribution over statistically independent variables captured by the minimal encoding can be efficiently captured using only $n_c$ continuous variables $\{w_i\}$. Each variable $w_i$ replaces $\langle \hat P_i^{1}\rangle_{\vec\theta} / \langle \hat P_i\rangle_{\vec\theta}$ in Eq.~\eqref{Eq:C1}, resulting in a non-convex quadratic optimization problem with continuous variables which can be solved using quadratic programming techniques~\cite{Wright2015}. 
This is in contrast to the number of parameters, $N_p = L \times n_q$, required in variational quantum circuits. From our numerical experiments shown in Fig.~\ref{Fig:CF_vs_L_Na1}, satisfactory results were obtained using $N_p = (L=4) \times (n_q = 4) > n_c = 8$ and $N_p = (L=16) \times (n_q = 6) > n_c = 32$ therefore suggesting classical approaches as more efficient routes. 
In the following section, we describe the methods for going beyond the limiting case of the minimal encoding where more sophisticated probability distributions can be captured by the quantum state through the use of additional ancilla qubits, therefore providing an opening for possible quantum advantages.


\section{Two-body correlations}

In this section, we show how two-body correlations between the classical variables of the QUBO problem can be introduced into the probability distribution captured by the quantum state.
These correlations refer to the conditional probability of one of the variables taking on a certain value given the value of another variable when sampling the classical solution from the probability distribution. 
We then describe how the particular topology of the different QUBO instances can influence the subset of correlated pairs to be encoded.
Specifically, when applied to a Max-Cut problem, we find that encoding only the correlations between pairs of variables that are connected within the graph leads to an improvement in the classical solutions obtained when compared to the minimal encoding approach.

\subsection{General encoding scheme}

We propose a general form of the quantum state that allows for the encoding of two-body correlations:
\begin{align} \label{Eq:Psi2}
	\ket {\psi_2(\vec\theta)} = \sum_{i,j}^{n_{\rm pair}} & \beta_{ij}(\vec\theta) [a_{ij}(\vec\theta)\ket{00}_a + b_{ij}(\vec\theta)\ket{10}_a  \\
	 & + c_{ij}(\vec\theta)\ket{01}_a + d_{ij}(\vec\theta)\ket{11}_a] \otimes \ket{\phi_{ij}}_r, \nonumber
\end{align}
where the register (ancilla) subspace now comprises $n_r = \log_2(n_{\rm pair})$ ($n_a=2$) qubits with $n_{\rm pair}$ being the number of two-body correlations encoded.
Similar to the minimal encoding scheme, each basis state $\ket{\phi_{ij}}_r$ of the register space acts as a pointer. However, this pointer now points to the index of a pair of classical variables $(x_i, x_j)$, as depicted in Fig.~\ref{Fig:Schema}(c). The associated two-qubit ancilla state encodes the bare probability for all pair values, e.g.~Pr$(x_i = 0, x_j = 0) = |a_{ij}|^2$, Pr$(x_i = 1, x_j = 0) = |b_{ij}|^2$ and so on.
This encoding allows one to produce probability distributions that is able to capture correlations beyond statistically independent variables. 
A similar encoding strategy has been considered to address the issue of limited connectivity in quantum annealing platforms, allowing to simulate all-to-all connectivity from only local interactions~\cite{Lechner2015}.

The form of Eq.~\eqref{Eq:Psi2} is general enough to allow correlations to be captured between either all pairs of variables or only a subset of these pairs. 
In certain cases, one might be able to infer a preferred subset of pairs to encode based on the specific topology of the problem, allowing for an important reduction in the number of qubits required.
In what follows, we highlight this point by comparing two general cases of frequently encountered QUBO models.

\subsubsection{Selective subsets for sparse matrices} \label{Sec:TBCMaxcut}

In QUBO instances where $\mathcal A$ is sparse, one might naturally expect that the most important correlations are those between the pairs of non-zero elements in $\mathcal A$.
One seminal instance of sparse QUBO models is the $d$-regular Max-Cut problem where $d \ll n_c$.
Each vertex on the corresponding graph is represented by a classical variable in $\vec x$ as depicted in Fig.~\ref{Fig:Schema} (b), and each edge by a non-zero off-diagonal element in $\mathcal A$.
The resulting matrix $\mathcal A$ has $d$ unit entries per row and column, and diagonal elements $\mathcal A_{i,i}=-d$. 
By selectively encoding only the $n_{\rm pair} = n_c\times d/2$ pairs between non-zero elements in $\mathcal A$ (i.e.~the edges), $n_q = \log_2(n_c\times d) + 1$ are required, which is only $\log_2(d)$ qubits more than the minimal encoding scheme.

Illustrating with an example, encoding the $12$ edges of the $3$-regular graph with $n_c=8$ shown in Fig.~\ref{Fig:Schema} (b) would require $n_r = 4$ register qubits. The pair $(x_1, x_2)$ could be mapped onto the basis state $\ket{\phi_{12}}_r \equiv \ket{0000}_r$, the pair $(x_1, x_7)$ on $\ket{\phi_{17}}_r \equiv \ket{0001}_r$ and so on until each edge is associated with a unique basis state.
In the later sections, we apply this selective encoding method to solve a $3$-regular Max-Cut problem with $n_c = 42$ number of variables using $n_q=8$ qubits, allowing us to surpass the performance of the minimal encoding scheme.

\subsubsection{Encoding all possible pairings for dense matrices}

For more extreme instances where $\mathcal A$ is dense, such as the randomly generated QUBO models used in the previous section, selecting a specific subset of two-body correlations becomes completely arbitrary.
The only unbiased approach then involves encoding all possible $n_{\rm pair} = n_c(n_c-1)/2$ pairs of classical variables, requiring the maximal number of qubits $n_q = \log_2[n_c(n_c-1)] + 1$. 
Using this method to encode the 28 edges in the fully connected graph shown Fig.~\ref{Fig:Schema} (a) would require $n_r = 5$ register qubits.
The mapping would proceed in a similar fashion as before, where the pair $(x_1, x_2)$ can be associated to $\ket{\psi_{12}}_r \equiv \ket{00000}_r$, $(x_1, x_3)$ to $\ket{\psi_{13}}_r \equiv \ket{00001}_r$ and so on.
Despite the ``unbiased'' choice of pairing the variables, capturing all possible two-body correlations for general dense QUBO problems is typically not an efficient use of quantum resource as we shall observe later.

\subsection{Averaging the probabilities and defining the cost function}

Interpreting the quantum state $\ket{\psi_2}$ in Eq.~\eqref{Eq:Psi2} as a distribution function Pr$(\vec x)$ over the ensemble of classical solutions $\vec x$ is not as straightforward as the minimal encoding case. 
To better understand this statement, let us first consider the limit where the ensemble of pairs $\{ (i,j)\}$ encoded would correspond to the set of edges in a 1-regular graph, also known as a perfect matching in graph theory and highlighted in Fig.~\ref{Fig:Schema} (b). 
In this case, each variable $x_i$ is paired with a single other variable $x_j$ and the probability to sample a solution $\vec x$ is uniquely defined as Pr$(\vec x) = \prod_{(i,j)} {\rm Pr}(x_i,x_j)$.
However, in the more general scenarios where at least one variable is included in more than one pair, the probability of sampling a solution $\vec x$ is not uniquely defined anymore.
For example, in the limit where all pairs are encoded, there are $N_{\rm pm}(n_c) = (n_c-1)!!$ ways of calculating Pr$(\vec x)$ with the possibility of vastly different results, where $N_{\rm pm}(n_c)$ is the number of perfect matchings in a fully connected graph.

In order to be able to define a cost function in the form of Eq.~\eqref{Eq:C1} that is well-behaved despite the non-uniqueness of Pr$(\vec x)$, we need to define averaged probabilities $\bar P^{i,j}_{\sigma_i,\sigma_j}$ of sampling $x_i = \sigma_i$ and $x_j = \sigma_j$ where $\sigma = \{0,1\}$ that takes into account the multiple ways of calculating Pr$(\vec x)$.
Doing so, we obtain the averaged probability of sampling $(x_i, x_j) = (1,1)$ from

\begin{align} \label{Eq:OffDiagMean}
	\bar P^{i,j}_{1,1} = \sum_{l\neq i,j}^{n_c}\sum_{k\neq i,j,l}^{n_c} & R_{ijkl}(\mathcal{G})(|c_{ki}|^2 + |d_{ki}|^2)(|c_{lj}|^2 + |d_{lj}|^2) \nonumber \\
	& + R_{ij}(\mathcal{G})|d_{ij}|^2,
\end{align}
where $c_{ij}$ and $d_{ij}$ are the amplitudes of the ancilla states given in Eq.~\eqref{Eq:Psi2} ($\vec\theta$ is kept implicit).
Here, $R_{ij}(\mathcal G) = N_{\rm pm}(\mathcal G - v_i - v_j)/ N_{\rm pm}(\mathcal G)$ is the ratio between the number of perfect matchings after subtracting the vertices $v_i$ and $v_j$ from the graph $\mathcal G$ to the total number of perfect matchings in $\mathcal G$. Similarly, $R_{ijkl}$ describes the same ratio but with $4$ vertices removed instead. The graph $\mathcal G$ is built by mapping each classical variable to a vertex and each pair encoded in $\ket{\psi_2}$ to an edge.
Expressions similar to Eq.~\eqref{Eq:OffDiagMean} for $\bar P^{i,j}_{0,0}$, $\bar P^{i,j}_{0,1}$ and $\bar P^{i,j}_{1,0}$ are derived in appendix~\ref{App:CostFunctions}.
Using the same approach, one can also derive the averaged probability of sampling $x_i = 1$, leading to
\begin{align} \label{Eq:MeanPri}
	\bar P^{i}_{1} = \sum_{k\neq i}^{n_c} R_{ik}(\mathcal{G})(|b_{ik}|^2 +|d_{ik}|^2),
\end{align}
where $b_{ij}$ is also defined in Eq.~\eqref{Eq:Psi2}.

In the limit where all possible pairs are encoded, these ratios are $R_{ij}(\mathcal G) = (n_c-3)!!/(n_c-1)!! = 1/(n_c-1)$ and $R_{ijkl}(\mathcal G) = R_{ij}(\mathcal G)/(n_c-3)$. However, in the case where only a subset of pairs are encoded, $R_{ijkl}(\mathcal G)$ depends on the vertices $\{i,j,k,l\}$ and is NP-hard to evaluate. 
One thus needs to resort to approximated ratios and our numerical experiments suggest that estimating $R_{ij}(\mathcal G) = 1/d$ and $R_{ijkl}(\mathcal G) = R_{ij}(\mathcal G)/(d-2)$ for a d-regular graph leads to adequate behaviour of the probabilities.

Having the averaged probabilities defined, one can propose a cost function of the form
\begin{align}
    C_2(\vec\theta)  = \sum_{i,j=1}^{n_c}& A_{ij}\bar P^{i,j}_{1,1}(\vec\theta)(1-\delta_{ij}) + \sum_{i=1}^{n_c} A_{ii}\bar P^{i}_{1}(\vec\theta).  
	\label{Eq:C2}
\end{align}
The key properties of Eq.~\eqref{Eq:C2} are similar to that of Eq.~\eqref{Eq:C1} in that (i) its global minimum corresponds unambiguously to the solution $\vec x$ that minimizes the QUBO problem, (ii) it can be estimated by a series of measurements solely in the computational basis, and (iii) it cannot be cast as a linear function of expectation values.

	
\begin{figure}
\center
\includegraphics[width=0.9\columnwidth]{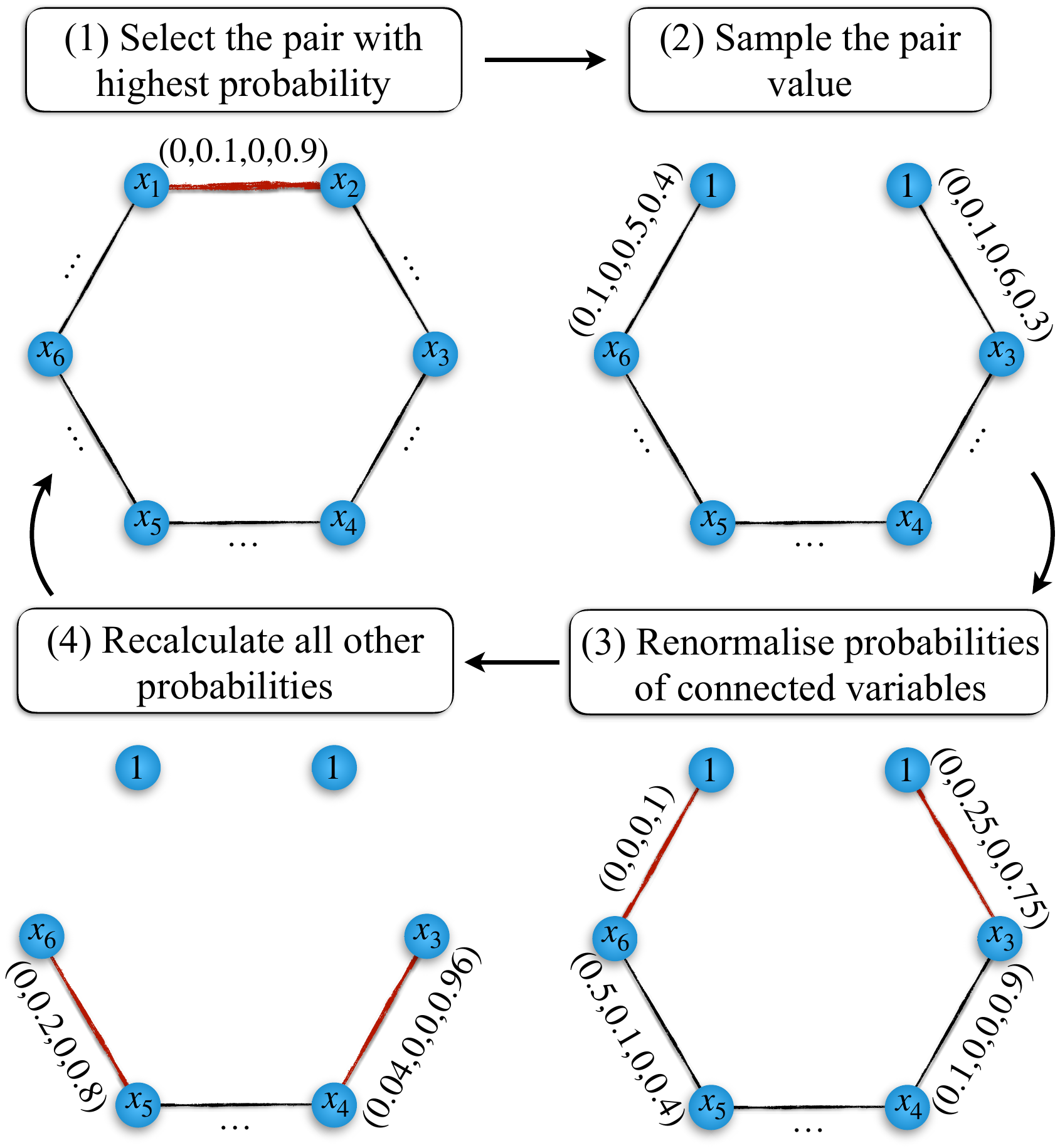}
\caption{
Schematic of the protocol to sample a solution from the ensemble of two-body correlations. The set of numbers displayed next to the edges between the vertices labeled $x_i$ and $x_j$ represents the probabilities $\bar P^{i,j} = (\bar P^{i,j}_{0,0}, \bar P^{i,j}_{1,0}, \bar P^{i,j}_{0,1}, \bar P^{i,j}_{1,1})$ with the convention $i<j$. 
In this example, assuming that $\bar P^{1,2}_{1,1} = 0.9$ has the largest probability in the entire graph, $(x_1, x_2)$ would be the first pair to be selected in step (1).
The remaining steps are outlined in the main text.
}
\label{Fig:Sampling}
\end{figure}


\subsection{Sampling the classical solution from the quantum state}

The form of $\ket{\psi_2(\vec\theta_{\rm opt})}$ provides some flexibility in how solutions can be sampled from it. 
In the following, we describe a sampling protocol that fully exploits the encoded correlations and we show a simple example in Fig.~\ref{Fig:Sampling}.

The procedure is as follows.
\begin{enumerate}

\item Select the pair $(i,j)$ with the most definite mean probabilities, i.e.~the pair where $\bar P^{i,j}_{\sigma_i, \sigma_j}$ of sampling $x_i = \sigma_i$ and $x_j = \sigma_j$ is the closest to unity. As an example, let us consider that the probability to sample $(x_1, x_2) = (1,1)$, $\bar P^{1,2}_{1,1} = 0.9$, is the largest of all mean probabilities, we select the pair $(1,2)$. 

\item Sample the value of the variables from the set of probabilities $\bar P^{i,j} = (\bar P^{i,j}_{0,0}, \bar P^{i,j}_{1,0}, \bar P^{i,j}_{0,1}, \bar P^{i,j}_{1,1})$.

\item Renormalize the probabilities of all variables connected to the pair evaluated in (2). 
Following the example, assume that $x_2 = 1$ has been sampled and is connected to the variable $x_3$, with probabilities $\bar P^{2,3} = (0, 0.1, 0.6, 0.3)$.
The probability of sampling $(x_2, x_3) = (0,1)$, $P^{2,3}_{0,1} = 0.6$, is now irrelevant given that $x_2 = 1$, leading to an updated probability of sampling $x_3 = 0$ of $0.25$ and Pr$(x_3 = 1) = 0.75$.

\item Adjust all remaining probabilities $\bar P^{k,l}$ consequently. Going back to the example, $x_3$ is now connected to $x_4$ where the probabilities were initially $\bar P^{3,4} = (0.1, 0, 0, 0.9)$. Since the probability to sample $x_3 = \{0,1\}$ have been modified as exemplified in (3),  $\bar P^{3,4}$ is re-evaluated to $(0.04, 0, 0, 0.96)$.

\item 
Repeat the steps from (1) until all variables have been assigned a value.

\end{enumerate}

Conceptually, this method allows for a finite propagation of correlations along the graph $\mathcal G$ during the sampling.
As an example, let us consider the case where correlations in the pairs $(x_i,x_k)$ and $(x_k,x_l)$ are explicitly encoded in $\ket{\psi_2}$ but not for the pair of variables $(x_i, x_l)$.
Using this sampling technique makes the probability of sampling $x_l = \{0,1\}$ change conditionally for the sampled value of $x_i$, therefore inducing correlations. 
We stress that these induced correlations are not captured in the optimization process, but only during sampling.

\subsection{Application to randomly generated QUBO instances versus a d-regular Max-Cut}

In this section, we present the results obtained after optimizing quantum states of the form of Eq.~\eqref{Eq:Psi2} using the cost function $C_2(\vec\theta)$ for two different instances of the QUBO model --- a 3-regular Max-Cut problem of $n_c = 42$ and a randomly generated matrix $\mathcal A$ of $n_c = 8$.

	
\begin{figure*}
\center
\includegraphics[width=0.95\textwidth]{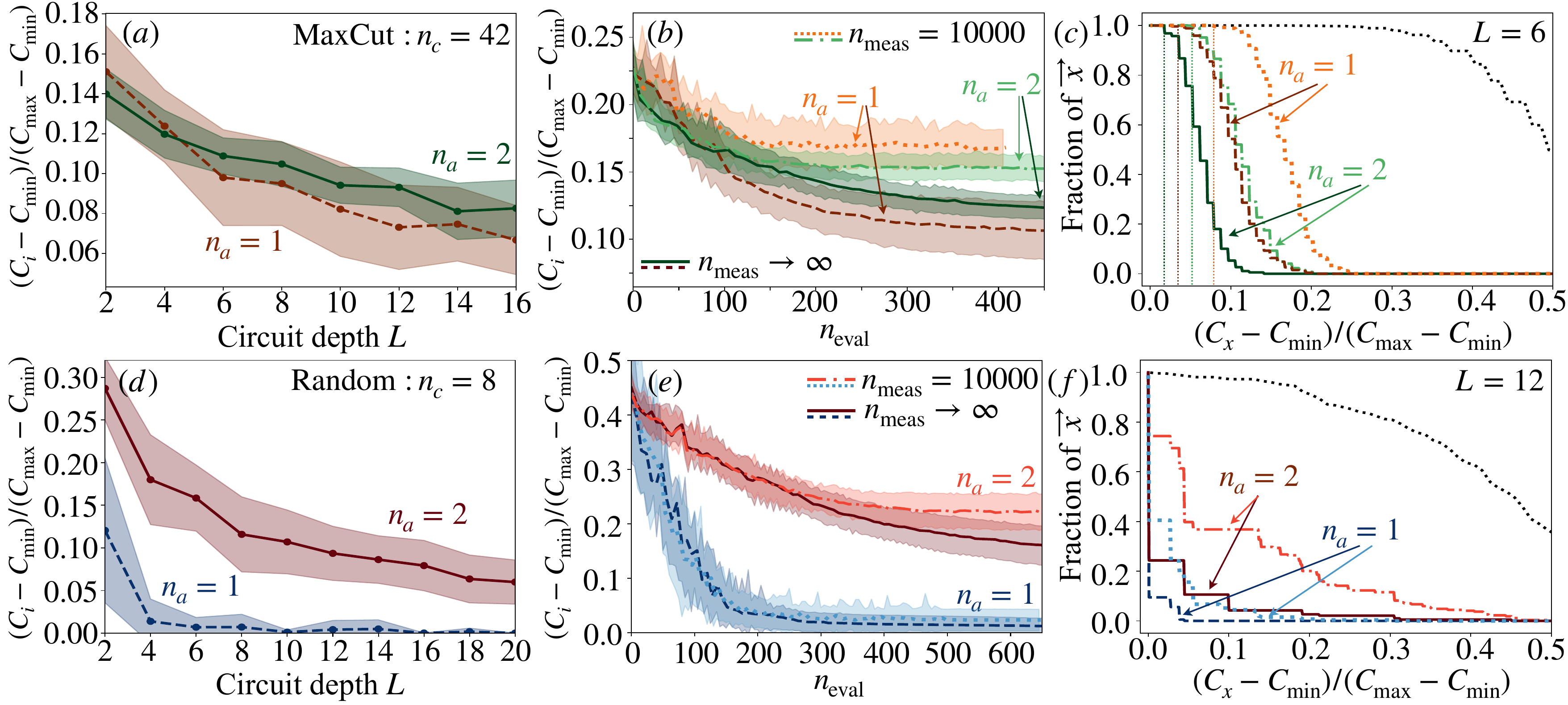}
\caption{
Comparison of results between $n_a = 1$ and $n_a = 2$ encoding schemes for a randomly generated matrix $\mathcal A$ of $n_c = 8$ classical variables with all $28$ possible pairings being encoded (top row) and a randomly generated 3-regular Max-Cut problem of $n_c = 42$ using selective encoding to encode only vertices joined by an edge (bottom row). The lighter and darker shaded lines show $n_{\rm meas} = 15000$ and $n_{\rm meas} \rightarrow \infty$ respectively. $n_q=8$ and $n_q=6$ ($n_q=7$ and $n_q=4$) qubits were used for the Max-Cut and randomly generated instances respectively for the $n_a=2$ ($n_a=1$) encoding scheme. (a),(d) Comparison of final cost function values $C_1(\vec \theta_{\rm opt})$ and $C_2(\vec \theta_{\rm opt})$ as a function of the circuit depth $L$. Points show the mean value over 30 starting points $\vec\theta_{\rm ini}$. Shaded regions represent one standard deviation from the mean. (b), (e) Comparison of $C_1(\vec \theta_{\rm opt})$ and $C_2(\vec \theta_{\rm opt})$ during the course of optimization. Mean number of evaluations for $n_{\rm meas} \rightarrow \infty$ extend beyond what is shown. (c), (f) Cumulative distribution of solutions drawn from the optimized quantum states $\ket{\psi_1(\vec \theta_{\rm opt})}$ of (b) and (e) respectively, sorted according to their energy. $10$ classical solutions, $\vec x$, were drawn from each optimized quantum state. The black dotted line shows the distribution of (c) all $2^{n_c} = 256$ solutions and (f) $4\times 10^8$ randomly generated solutions.
}
\label{Fig:CF_vs_L_Na2}
\end{figure*}


\subsubsection{Selective encoding for a 3-regular Max-Cut problem}

To demonstrate the effectiveness of capturing correlations, we apply our encoding scheme for $n_a=2$ to a randomly generated 3-regular Max-Cut problem with $n_c = 42$ vertices and $63$ edges. In this example, selective encoding was used to only encode correlations between classical variables that are connected by an edge, requiring $n_q = \log_2(63) + 2 \lesssim 8$ qubits. By contrast, encoding all of the $861$ possible pairs would require $12$ qubits.

Using the same hardware-efficient circuit shown in Fig.~\ref{Fig:Circuit}, we apply the optimization protocol described in Sec.~\ref{Sec:Training1q}.
In Fig.~\ref{Fig:CF_vs_L_Na2} (a), we show the final cost function $C_2(\vec\theta_{\rm opt})$ as a function of the circuit depth $L$ in the limit of $n_{\rm meas} \rightarrow \infty$.
We compare to optimization results for the same problem using the minimal encoding scheme $n_a = 1$.
Panel (b) shows the differences in the optimization process between the $n_a=1$ and $n_a=2$ encoding schemes for $L=6$ for $n_{\rm meas} \rightarrow \infty$ and $n_{\rm meas} = 10^4$.
While $C_1$ and $C_2$ are both depicted in the same figure to demonstrate their respective performance, we stress that they are different quantities and might lead to substantial differences in the quality of the solutions sampled despite their comparable values. 
This discrepancy is further accentuated given the fundamentally different sampling protocols.

The distribution of solutions drawn from $\ket{\psi_2}$ show a substantial improvement in quality over the solutions obtained from $\ket{\psi_1}$, as depicted in Fig.~\ref{Fig:CF_vs_L_Na2} (c). 
Importantly, as we show in Appendix \ref{App:Noise}, this improvement over the minimal encoding is preserved and even amplified in presence of experimental imperfections. Intuitively, this enhanced robustness to noise could be the results of the information redundancy encoded when capturing two-body correlations, a characteristic reminiscent of the general idea of error-correction.
The use of selective encoding has thus allowed us to produce better quality solutions through a combination of encoding only the subset of two-body correlations that are expected to be the most relevant and reducing the complexity of the cost function $C_2(\vec\theta_{\rm opt})$.

\subsubsection{Encoding all pairs for randomly generated QUBO instances}
We conclude the results by revisiting the matrix $\mathcal A$ with $n_c = 8$ consisting of elements drawn from a continuous uniform probability distribution. 
In this instance, all $28$ possible pairings between the $8$ classical variables are encoded, requiring a total of $n_q = \log_2(28) + 2 \lesssim 7$ qubits. 

The results are shown in Fig.~\ref{Fig:CF_vs_L_Na2} (d)--(f) and compared to the results previously obtained in the minimal encoding scheme. 
Most importantly, panel (c) shows that solutions sampled from the statistically independent distribution function encoded in $\ket{\psi_1}$ are of better quality than those sampled from $\ket{\psi_2}$. 
These results strongly suggest that encoding all pairs is not an efficient use of quantum resources and can lead to a poorer performance during optimization as well as poorer quality solutions obtained from the final state.

Intuitively, this efficiency loss can be attributed to the use of a much larger Hilbert space to encode highly redundant, and possibly contradictory, information about classical correlations. This suggests that there is a balance to reach regarding the subset of pairs to be encoded and the resources required to do so; a more detailed discussion of the general case for any choice of $n_a$ is presented in the following section.


\section{Generalization to multi-body correlations} \label{generalization}

Now that we have described in detail a framework to make the first step beyond statistically independent classical variables and encode two-body correlations, generalizing the idea to encoding any set of $n_a$-body correlations is straightforward.
Consider a variational quantum state of the form:
\begin{equation}
    \ket{\psi_a(\vec\theta)} = \sum_i \beta_i(\vec\theta)\ket{\varphi_i(\vec\theta)}_a\ket{\Phi_i}_r,
\end{equation}
where the ancilla state $\ket{\varphi_i(\vec\theta)}_a$ is composed of $n_a$ qubits and is associated with a register state $\ket{\Phi_i}_r$ that points to a specified group $i$ of $n_a$ classical variables.
In light of the previous section, whether $\ket{\psi_a(\vec\theta)}$ can be efficiently optimized to solve a QUBO problem strongly depends on the choice of the encoded groups of $n_a$ classical variables.

One of the simplest mapping strategies consists of encoding a selected set of $n_c/n_a$ independent groups of $n_a$ variables, i.e.~where no variable is part of more than one group.
The number of qubits needed for this,
\begin{equation}
    N_{\rm ind}(n_a) = \log_2(n_c/n_a) + n_a, \label{Eq:independent_na}
\end{equation}
increases monotonically until the complete encoding threshold where $n_a=n_c$.
In this strategy, there is a one-to-one correspondence between each of the $n_c/n_a$ subgroup of $n_a$ classical variables and a unique basis state of the $n_r = \log_2(n_c/n_a)$ register qubits.
The quantum state $\ket{\varphi_i(\vec\theta)}_a$ of the $n_a$ ancilla qubits associated with the $i^{\rm th}$ subgroup encodes a distribution function that can capture all correlations among the variables of this subgroup.
The optimization protocol can be interpreted as partitioning the QUBO problem into subgroups and simultaneously solving each of them using the complete encoding.  
This choice of mapping is one that is arbitrary as there is no fixed structure as to how the variables should be grouped. However, the minimal use of quantum resources might make this a desirable choice in certain situations.

Another strategy would be to encode all $\frac{n_c!}{n_a!(n_c-n_a)!}$ groups of $n_a$ variables, which is the generalization of encoding all possible pairs for $n_a=2$. This requires 
\begin{equation}
    N_{\rm all}(n_a) = \log_2\left(\frac{n_c!}{n_a!(n_c-n_a)!}\right) + n_a
\end{equation}
qubits, which is a non-monotonic function of $n_a$ and can substantially exceed the total number of qubits required for the complete encoding, showing an inefficient use of quantum resource. 

In between these two extremes are multiple mapping options and whether any of these encoding schemes can efficiently exploit the dominant correlations within a specific family of QUBO models is of great interest. For example, one could imagine encoding an ensemble of $(d+1)$-body correlations that follows the specific topology of a $d$-regular Max-Cut problem. In this case, each classical variable within the $d$-regular graph forms a group of $d+1$ elements. Encoding all of those $n_c$ groups into a quantum state would require
\begin{equation}
    N_{\rm reg}(n_a) = \log_2(n_c) + n_a + 1,
\end{equation}
qubits, where $d=n_a$.
For $n_a\rightarrow (n_c-1)$, i.e.~a fully connected graph, the number of qubits exceeds the threshold $n_q=n_c$ by $\log_2(n_c)$.

To investigate the resources required to reproduce these probability distributions classically, we consider the simplest encoding strategy described in Eq.~\eqref{Eq:independent_na} where the classical variables are grouped into distinct subgroups.
In this scenario, the cost function to minimize is a direct generalization of Eq.~\eqref{Eq:C1}, i.e.~it can still be cast as a quadratic optimization problem over continuous variables.
This time however, capturing all correlations encoded in the quantum state would require $2^{n_a}$ classical variables for each of the $2^{n_r}$ subsystems, leading to a total of $2^{n_q}$ total variables, as expected.
In contrast, by using a variational quantum circuit, the number of variational parameters involved during optimization remains $N_p = L \times n_q$.
While it is expected for $L$ to scale with the number of qubits involved, the exact nature of this scaling is still an open question. Anything sub exponential, which can be expected from previous analysis in the context of random quantum circuits~\cite{Brandao2016}, could lead to quantum advantages. 
One indication favouring this sub exponential depth can be seen in the context of random circuits where an ensemble of random unitaries with approximate t-design properties can be produced with polynomial circuit depth. 
Because a logarithmic compression in the number of qubits is unlikely to bring about any computational advantages (cf.~Section \ref{Sec:classicalsim}), the crossover between the minimal encoding and the complete encoding, where the compression in number of qubits is polynomial, is of great interest and needs to be further studied.


\section{Conclusion}

In this work, we have proposed and analysed a systematic encoding scheme for variational quantum algorithms that allows one to capture increasing amount of correlations between classical variables in optimization problems. 
We first detailed the implementation of the minimal encoding scheme, using only $n_q = \log_2(n_c) + 1$ qubits to solve a QUBO model of size $n_c$. 
This significant reduction in qubits allowed us to tackle randomly generated problem instances of size $n_c = 8, 32$ and $64$ using only $n_q = 4, 6$ and $7$ qubits respectively. 
Our numerical solutions was able to find suitable high quality solutions using resources compatible with NISQ devices despite the inability to capture any correlations between the classical variables. 
The use of a hardware efficient parameterized circuit allowed us to reduce the number of gates required during implementation. Implementing QAOA according to recent state-of-the-art experiments in performed in Ref.~\cite{Arute2020} would require $612$ gates for $p=3$ when applied to an $n_c=8$ variable problem. However, superior solutions can be obtained using our minimal encoding scheme with as little as $42$ gates and $n_q=4$ qubits.

We also detailed encoding protocols that allow for two-body correlations to be captured between the classical variables. The number of qubits required scales logarithmically with the number of pairs encoded and we showed that exploiting the topology of the QUBO instance is essential for efficient optimization of the quantum state. 
By applying the two-body correlation encoding to a Max-Cut problem of 42 vertices, we were able to obtain better performance compared to the minimal encoding scheme.

The focus of this work was primarily on the encoding schemes outlined in the main text and was not intended as a thorough investigation of the most efficient optimization protocols. We believe that the results presented can still be improved upon substantially. One possible area for exploration could be finding an ansatz that would result in a smoother cost function landscape with shallower circuits. More adapted classical optimization methods may also bring significant improvements in the optimization process as it was found that a considerable fraction of optimization runs got stuck in local minimas \cite{sung2020, lavrijsen2020}. 
Improvement on that front may also substantially decrease the number of measurements required to reach comparable quality of solutions.
Further avenues to explore would be whether generalizations to larger $n_a$-body correlations can be efficiently optimized and whether alternative ways of capturing correlations for dense problem instances can be found. 
More importantly, we wish to investigate the intermediate encoding schemes beyond the limit of classical intractability where quantum algorithms may outperform classical approaches.

\section{Acknowledgments}

This research is supported by the National Research Foundation, Prime Minister’s Office, Singapore and the Ministry of Education, Singapore under the Research Centres of Excellence programme. It was also partially funded by Polisimulator project co-financed by Greece and the EU Regional Development Fund, the European Research Council under the European Union’s Seventh Framework Programme (FP7/2007-2013).


\appendix

\section{Derivation of the cost functions} 
\label{App:CostFunctions}

In all of the encoding schemes outlined in the main text, the quantum state $\ket{\psi(\vec\theta)}$ captures a probability distribution over all $2^{n_c}$ classical solutions. In this context, we generalize the QUBO cost function, $C_{x} = \vec x^\intercal A \vec x$, as a sum over all possible solutions weighted by their respective probability to be sampled, i.e.
\begin{align}
	\mathcal C = & \sum_{\{\vec x\}} \vec x^\intercal \mathcal A \, \vec x \, {\rm Pr} (\vec x), \nonumber \\
	= & \sum_{\{\vec x\}} \sum_{i,j=1}^{n_c} x_i \mathcal A_{ij} x_j {\rm Pr} (\vec x), \nonumber \\
	= & \sum_{i,j=1}^{n_c} \mathcal A_{ij} \sum_{\{\vec x | x_i = x_j = 1\}} {\rm Pr} (\vec x). \label{Eq:cf_deriv_3}
\end{align}
Here, $\{\vec x\}$ represents the ensemble of all the $2^{n_c}$ possible solutions $\vec x$ while $\{\vec x | x_i = x_j = 1\}$ represents only the subset of the $2^{n_c-2}$ solutions where the $i^{\rm th}$ and $j^{\rm th}$ variables in $\vec x$ are $x_i = x_j = 1$.
To obtain the third line, we have explicitly used the fact that only variables with values equal to $1$ contributes to the cost function. 
In what follows, we present in more details the following steps that lead to Eqs.~\eqref{Eq:C1} and \eqref{Eq:C2} of the main text and provide further discussions about their properties.

\subsection{Minimal encoding}

In the minimal encoding, the state $\ket{\psi_1(\vec\theta)}$ describes statistically independent classical variables where the probability of sampling $\vec x$ is Pr$(\vec{x}) = \prod_{i=1}^{n_c}{\rm Pr}(x_i)$. 
In this case,
\begin{equation}
    \sum_{\{\vec x | x_i = x_j = 1\}} {\rm Pr} (\vec x) = \left\lbrace\begin{matrix} {\rm Pr}(x_i = 1){\rm Pr}(x_j = 1) & {\rm if} & i\neq j \\
    {\rm Pr}(x_i = 1) & {\rm if} & i = j
    \end{matrix}\right.,
\end{equation}
which, in terms of the quantum state amplitudes, reads
\begin{equation}
    \sum_{\{\vec x | x_i = x_j = 1\}} {\rm Pr} (\vec x) = \left\lbrace\begin{matrix} |b_i(\vec\theta)|^2|b_j(\vec\theta)|^2 & {\rm if} & i\neq j \\
    |b_i(\vec\theta)|^2 & {\rm if} & i = j
    \end{matrix}\right..
\end{equation}
By substituting these results into Eq.~\eqref{Eq:cf_deriv_3}, one gets
\begin{equation}
    C_1(\vec\theta) = \sum_{i,j=1}^{n_c} \mathcal A_{ij} |b_i(\vec\theta)|^2|b_j(\vec\theta)|^2(1 - \delta_{ij}) + \sum_{i=1}^{n_c} \mathcal A_{ii}|b_i(\vec\theta)|^2.
\end{equation}
The final form presented in Eq.~\eqref{Eq:C1} of the main text is obtained by expressing the probabilities $|b_i(\vec\theta)|^2 = \langle \hat P_i^{1}\rangle_{\vec\theta}/\langle \hat P_i\rangle_{\vec\theta}$ in terms of the projectors $\hat P_i$ and $\hat P_i^{1}$ (defined in the main text).

\subsection{Two-body correlations}

In the case where the variational quantum state $\ket{\psi_2(\vec\theta)}$ encodes a given set of two-body correlations, evaluating Eq.~\eqref{Eq:cf_deriv_3} is not as straightforward as in the minimal encoding. 
This is due to the multiple ways of evaluating the probability Pr$(\vec x)$ of sampling a solution $\vec x$, each of which capable of producing very different results.
More precisely, for $\vec x = (\sigma_1, \sigma_2, \dots, \sigma_{n_c})$, Pr$(\vec x) = \prod_{\{(i,j)\}}{\rm Pr}(x_i=\sigma_i| x_j = \sigma_j)$ where ${\rm Pr}(x_i=\sigma_i| x_j = \sigma_j)$ represents the conditional probability to sample $x_i = \sigma_i$ given $x_j = \sigma_j$. Here the ensemble $\{(i,j)\}$ represents a set of independent encoded pairs where no variables are repeated, i.e.~a perfect matching. Consequently, there are as many ways to evaluate Pr$(\vec x)$ as there are perfect matchings $N_{\rm pm}(\mathcal G)$ in the graph $\mathcal G$, corresponding to the encoded pairs in $\ket{\psi_2(\vec\theta)}$.

To evaluate Eq.~\eqref{Eq:cf_deriv_3}, we average over all possible ways of evaluating Pr$(\vec x)$, denoted by $\{ {\rm Pr}(\vec x)\}$, and define the mean probabilities
\begin{align}
    \bar{P}^{i,j}_{1,1} \equiv \frac{1}{N_{\rm pm}(\mathcal{G})} \sum_{\{{\rm Pr}(\vec x)\}} \sum_{\{\vec x | x_i = x_j = 1\}} {\rm Pr} (\vec x),
\end{align}
for $i\neq j$. The mean probability to sample a single variable $x_i = 1$, $\bar{P}^{i}_{1}$, is given by the same above definition with $i=j$.
There are two distinct scenarios that one can encounter while averaging over all possible perfect matchings corresponding to $x_i=x_j=1$ in $\mathcal G$. The first is when the perfect matching contains an edge connecting $x_i$ and $x_j$. There are $N_{\rm pm}(\mathcal G_{ij})$ of such instances, where $\mathcal G_{ij}$ is the graph obtained by subtracting the two vertices $i$ and $j$. For each of these instances, the conditional probability ${\rm Pr}(x_i=1| x_j = 1) = |d_{ij}(\vec\theta)|^2$ is directly encoded in the quantum state (see Eq.~\eqref{Eq:Psi2} of the main text). 
The second scenario occurs when the perfect matching does not include an edge connecting the vertices $i$ and $j$ to each other but instead to other vertices $k$ and $l$. 
These cases appear within a subset of $N_{\rm pm}(\mathcal G_{ijkl})$ perfect matching instances, where $\mathcal G_{ijkl}$ is the graph obtained by subtracting the vertices $i,j,k$ and $l$.
In these scenarios, the conditional probability ${\rm Pr}(x_i=1| x_j = 1)$ is not directly encoded in the quantum state and has to be inferred from ${\rm Pr}(x_i=1| x_j = 1) = {\rm Pr}(x_k=0,1| x_i = 1){\rm Pr}(x_l=0,1| x_j = 1) = (|c_{ki}|^2 + |d_{ki}|^2)$, where ${\rm Pr}(x_k=0,1| x_i = 1)$ is the conditional probability of having $x_k=0$ or $x_k=1$ given $x_i = 1$. 

Considering these contributions, we obtain the following mean conditional probabilities:
\begin{widetext}
\begin{align} 
    \bar{P}^{i,j}_{1,1} =& \; \frac{N_{\rm pm}(\mathcal{G}_{ijkl})}{N_{\rm pm}(\mathcal{G})} \sum_{l\neq i,j}^{n_c}\sum_{k\neq i,j,l}^{n_c} {\rm Pr}(x_k = 0,1| x_i = 1) \: {\rm Pr}(x_l = 0,1|x_j = 1) + \frac{N_{\rm pm}(\mathcal{G}_{ij})}{N_{\rm pm}(\mathcal{G})}  {\rm Pr}(x_i=1|x_j=1), \nonumber\\
    =&
    \;\frac{N_{\rm pm}(\mathcal{G}_{ijkl})}{N_{\rm pm}(\mathcal{G})} \sum_{l\neq i,j}^{n_c}\sum_{k\neq i,j,l}^{n_c} \left[ \left(|c_{ki}|^2 + |d_{ki}|^2 \right) \left( |c_{lj}|^2 + |d_{lj}|^2 \right)\right] 
    + \frac{N_{\rm pm}(\mathcal{G}_{ij})}{N_{\rm pm}(\mathcal{G})}  |d_{ij}|^2, \label{Eq:cf2_offdiagderiv2}
\end{align}
for ($i\neq j$), and 
\begin{align} 
    \bar{P}^{i}_{1,1}=& \;\frac{N_{\rm pm}(\mathcal{G}_{ik})}{N_{\rm pm}(\mathcal{G})} \sum_{i\neq k}^{n_c} {\rm Pr}(x_k=0,1|x_i=1), \nonumber\\
    =& \; \frac{N_{\rm pm}(\mathcal{G}_{ik})}{N_{\rm pm}(\mathcal{G})} \sum_{i\neq k}^{n_c}  \left(|c_{ki}|^2 + |d_{ki}|^2 \right), 
    \label{Eq:cf2_diagderiv2}
\end{align}
for $i=j$.
\end{widetext}
The cost function in Eq.~\eqref{Eq:cf_deriv_3} thus adopts the final form
\begin{equation}
    C_2 =  \sum_{i,j=1}^{n_c} \mathcal A_{ij} \bar{P}^{i,j}_{1,1}(1 - \delta_{ij}) + \sum_{i=1}^{n_c} \mathcal A_{ii}\bar{P}^{i}_{1,1},
\end{equation}
as in Eq.~\eqref{Eq:C2} of the main text.

This averaging ensures a well behaved cost function where the quantum state which minimizes this cost function gives the unit probability of sampling the exact solution which minimizes the QUBO problem. 
The drawback of this method is the partial ``washing out'' of the encoded correlations as it can be seen by the first term (second scenario) in Eq.~\eqref{Eq:cf2_offdiagderiv2} which adopts the form of two statistically independent variables.

\newpage 

Following the steps outlined above, the following averaged probabilities can also be derived:
\begin{align}
    \bar{P}^{i,j}_{0,1} =& \sum_{l\neq i,j}^{n_c}\sum_{k\neq i,j,l}^{n_c} R_{ijkl}(\mathcal{G})(|a_{ki}|^2 + |b_{ki}|^2)(|c_{lj}|^2 + |d_{lj}|^2) \nonumber\\
	& + R_{ij}(\mathcal{G})|c_{ij}|^2, \\
    \bar{P}^{i,j}_{1,0} =& \sum_{l\neq i,j}^{n_c}\sum_{k\neq i,j,l}^{n_c}  R_{ijkl}(\mathcal{G})(|c_{ki}|^2 + |d_{ki}|^2)(|a_{lj}|^2 + |b_{lj}|^2) \nonumber\\
	& + R_{ij}(\mathcal{G})|b_{ij}|^2, \\
	\bar{P}^{i,j}_{0,0} =& 1 - \left[ \bar{P}^{i,j}_{0,1} - \bar{P}^{i,j}_{1,0} - \bar{P}^{i,j}_{1,1} \right].
\end{align}


\begin{figure*}
\center
\includegraphics[width=0.85\textwidth]{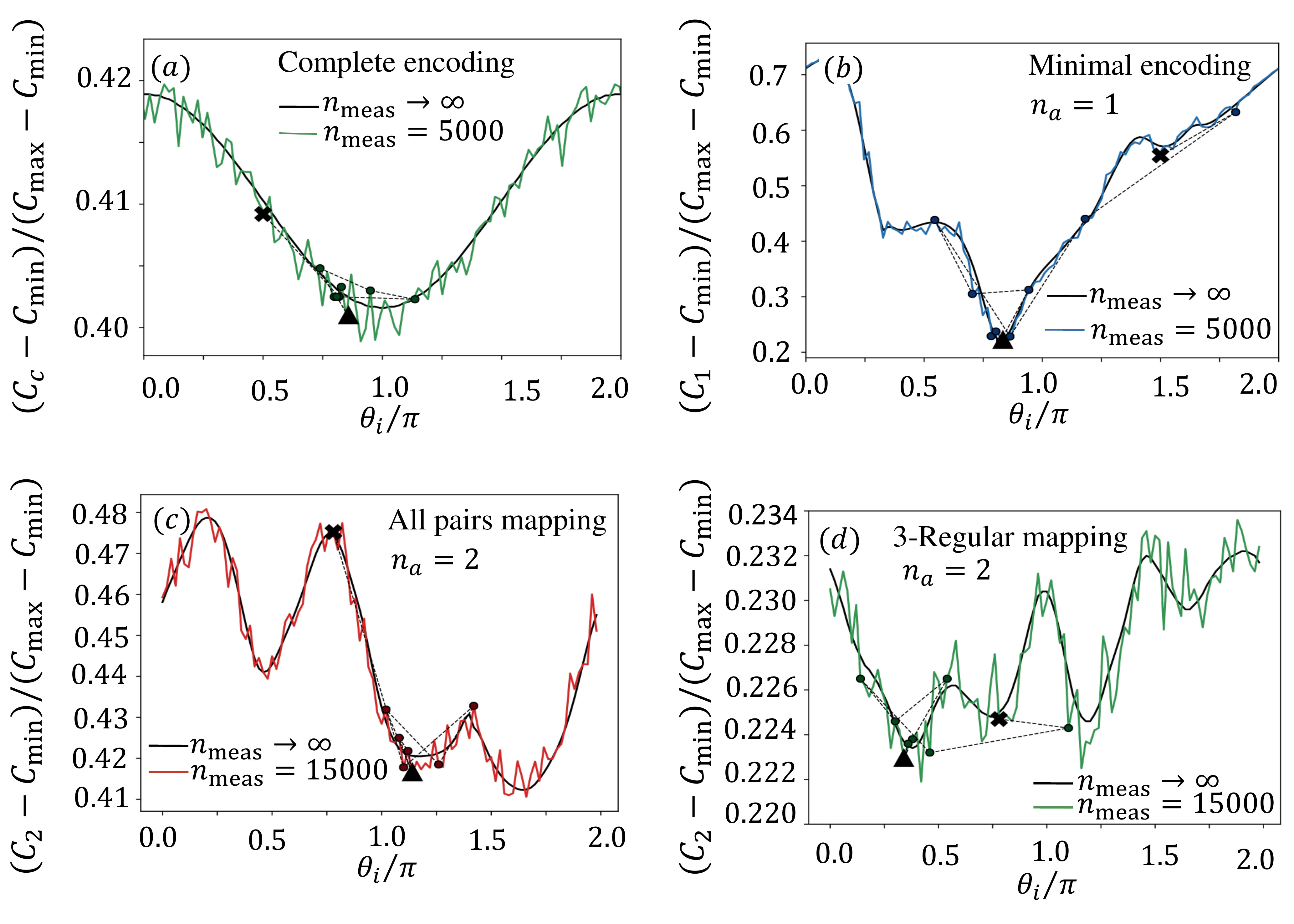}
\caption{
Cost function landscape as a function of a randomly chosen $\theta_i$. Solid line shows the expectation value in the limit $n_{\rm meas} \rightarrow \infty$, compared with the simulated value obtained from finite $n_{\rm meas}$. The dotted lines from the cross to the triangle show the path taken by the COBYLA optimizer to find the optimal $\theta_i$. 
(a)--(b) Complete and minimal encoding cost functions, $C_{\rm cp}(\vec \theta)$ and $C_1(\vec \theta)$, for a randomly generated $\mathcal A$ with $n_c=8$. Simulated values obtained using $n_{\rm meas}=5000$ and circuit depth $L=4$. $n_q=8$ qubits were used for the complete encoding compared to $n_q=4$ in the minimal encoding.
(c) $C_2(\vec \theta)$ for a randomly generated $\mathcal A$ with $n_c=8$ using $n_a=2$ encoding scheme. All possible two-body correlations were encoded using $n_q = 7$ qubits. Simulated values obtained using $n_{\rm meas}=15000$ and circuit depth $L=12$. (d) $C_2(\vec \theta)$ for a $3$-regular Max-Cut problem with $n_c=42$ classical variables using $n_a=2$ encoding scheme. Selective encoding with $n_q = 8$ qubits was used. Simulated values obtained using $n_{\rm meas}=15000$ and circuit depth $L=6$.
}
\label{Fig:CFlandscape}
\end{figure*}


\section{The cost function landscape}

The cost functions $C_1(\vec\theta)$ and $C_2(\vec\theta)$, described by Eq.~\eqref{Eq:C1} and Eq.~\eqref{Eq:C2} of the main text, are nonlinear combinations of expectation values.
This form leads to very different behaviours as a function of $\vec\theta$ when compared to the linear cost function $C_{\textrm{cp}}(\vec\theta)$ derived in the complete encoding limit [cf.~Eq.~\eqref{Eq:CF_FullEnc} of the main text].

These differences are depicted in Fig.~\ref{Fig:CFlandscape}, where $C_{\textrm{cp}}(\vec\theta)$, $C_1(\vec\theta)$ and $C_2(\vec\theta)$ are plotted as a function of a single parameter $\theta_i$ with all other rotation angles being fixed at random values.
The $\mathcal A$ matrix used in Fig.~\ref{Fig:CFlandscape} (a)--(c) is the same randomly generated $n_c=8$ matrix used in Section \ref{Sec:Training1q} of the main text,
while the same $\mathcal A$ matrix describing the $n_c=42$ $3$-regular Max-Cut in Section \ref{Sec:TBCMaxcut} was used in panel (d).
The circuit used to obtain the landscape of $C_{\textrm{cp}}(\theta_i)$ consists of a single layer of $R_Y(\theta)$ applied in parallel to all qubits.
This circuit was chosen as it consists of only single-qubit rotations with no entangling gates. The resulting quantum state can therefore only describe probability distributions of statistically independent classical variables in the complete encoding, and is equally expressible as $\ket{\psi_1(\vec \theta)}$ in the minimal encoding.

For deep circuits and linear cost functions, Ref.~\cite{McClean2018} predicts the existence of barren plateaus for $2$-design quantum circuits $\hat U(\vec\theta)$. 
Interestingly, the non linear forms of $C_1(\vec\theta)$ and $C_2(\vec\theta)$ do not fulfil the necessary conditions underlying the proof derived in Ref.~\cite{McClean2018}.
Consequently, we expect that a more constrained condition of a $t$-design quantum circuit, where $t>2$, would be necessary to demonstrate the existence of these barren plateaus. 
In addition, for cost functions comprising of a linear combination of a \textit{Poly}$(n_q)$ number of global observables, Ref.~\cite{cerezo2020} predicts the existence of barren plateaus even for shallow circuits. 
Despite the fact that each observable considered in this work is a projector, i.e.~global operator, the nonlinearity of $C_1(\vec\theta)$ and $C_2(\vec\theta)$ combined with the $\mathcal O(2^{n_q})$ number of terms involved also do not fulfil the necessary conditions for the proof in Ref.~\cite{cerezo2020}.
A more thorough investigation of the barren plateaus for nonlinear cost function is left for future work.

\section{Effects of noise}
\label{App:Noise}

\begin{figure}
\center
\includegraphics[width=0.9\columnwidth]{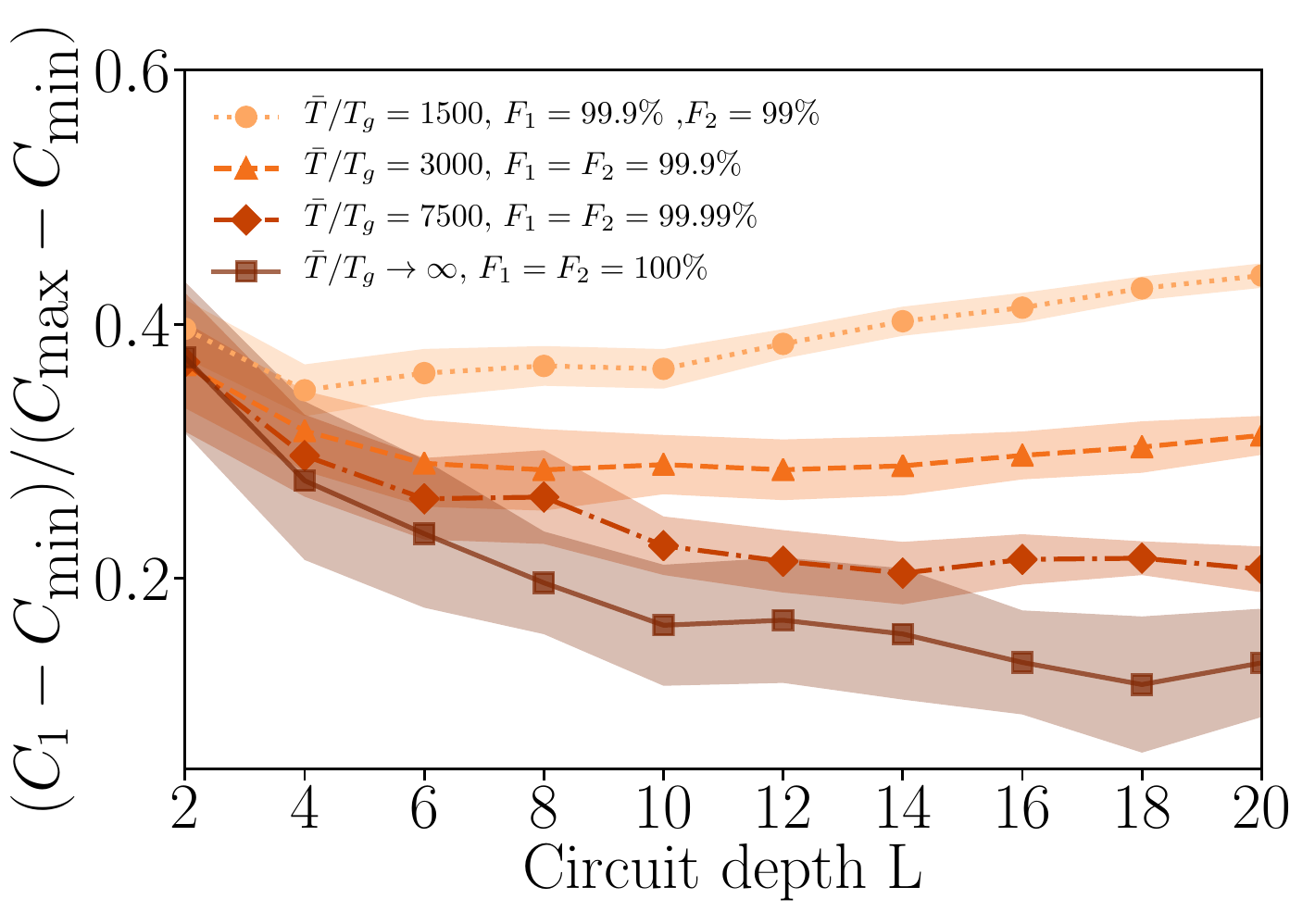}
\caption{
Optimized cost function value using the minimal encoding scheme as a function of circuit depth $L$ under the effects of various noise levels for $n_c=32$ classical variables and $n_{\textrm{meas}} = 5 \times 10^4$. 
$T_1/T_g$ and $T_2/T_g$ values for each qubit in the circuit were drawn from a normal distribution with a mean of $\bar T_1 = \bar T_2 \equiv \bar T$ and standard deviation $\sigma = \bar T/20$ obtained from surveying available quantum devices on the IBMQ cloud-based quantum computing platform~\cite{ibmq}. Here $T_g$ is the average single-qubit gate time and we used $T_{\rm CNOT}/T_g = 6$ and $T_{\rm meas}/T_g = 30$ where $T_{\rm CNOT}$ and $T_{\rm meas}$ are the average time for performing a CNOT gate and a measurement respectively.
$F_1$ and $F_2$ are the gate fidelities for single-qubit and two-qubit operations respectively. We used a readout error of $1\%$ for  all curves except the darkest plain line for which we used $0\%$. 
All other parameters are identical to Fig.~\ref{Fig:CF_vs_L_Na1}.}
\label{Fig:CFwithnoise} 
\end{figure}

In this section, we investigate the performance of our encoding scheme under the effects of a noise model consisting of thermal relaxation errors, imperfect gate fidelities, and readout errors. Thermal relaxation and decoherence can be characterized by the relaxation constants $T_1$ and $T_2$ (distinct from $T_2^*$) respectively. Given a single-qubit density matrix $\rho$, the effects of thermal processes can be simulated by transforming $\rho$ after a time evolution $t$ as

\begin{equation}
    \rho(t) \rightarrow \frac{1}{\rho_{00} + \rho_{11}e^{-t/T_1}}\begin{bmatrix} 
    \rho_{00} & \rho_{01}e^{-t/T_2} \\
    \rho_{10}e^{-t/T_2} & \rho_{11}e^{-t/T_1}
    \end{bmatrix}
    .
\end{equation}

Gate errors are implemented via a depolarization channel that affects each qubit as it undergoes a gate operation. On top of its intended operation, a gate with error $\lambda$ has an additional effect on $\rho$ according to

\begin{equation}
    \rho \rightarrow (1-\lambda) \rho + \frac{\lambda}{2^{n_q}} I
\end{equation}

where $I$ is the identity matrix representing the maximally mixed state. 
Readout error is the probability of obtaining an incorrect value of the qubit during measurement, i.e. reading a $\ket{0}$ when the qubit is in the $\ket{1}$ state and vice versa. In experimental quantum platforms, the magnitude for the errors above can differ between qubits, and we implement this model by assigning the qubits values drawn from a normal distribution characterized by a mean and standard deviation for each type of error.

Figure \ref{Fig:CFwithnoise} shows a comparison in the performance of the minimal encoding as a function of circuit depth $L$ for different levels of noise.
Each data point is obtained by performing the entire optimization protocol in the presence of all the noise sources described above. The lightest orange dashed line (circle data points) shows the results using noise levels characteristic of existing state-of-the-art hardware \cite{ibmq}. For comparison, we also consider more optimistic values that can be expected in the upcoming generations of NISQ devices, shown by the darker lines. The simulation for the triangle data points was achieved using a $2 \times$ increase in $\bar{T}$ and an increase in 2-qubit gate fidelity from $F_2 = 99\%$ to $F_2 = 99.9\%$ compared to the circle data points. The diamond points were obtained using a $5 \times$ increase in $\bar{T}$ from the circle data points, and an increase in single and 2-qubit gate fidelities to $F_1 = F_2 = 99.99\%$. Mean readout errors were kept unchanged at $1\%$ for all the noisy simulations
For comparison, the darkest line (square data points) corresponds to noise-free simulations, i.e.~$T=\infty$, $F_1 = F_2 = 1$ and perfect readout but with finite $n_{\textrm{meas}}$. The result of these optimistic noise levels can be expected from a direct improvement in hardware implementation or from applying additional error-mitigation techniques \cite{endo2020hybrid}.

\begin{figure}
\center
\includegraphics[width=0.95\columnwidth]{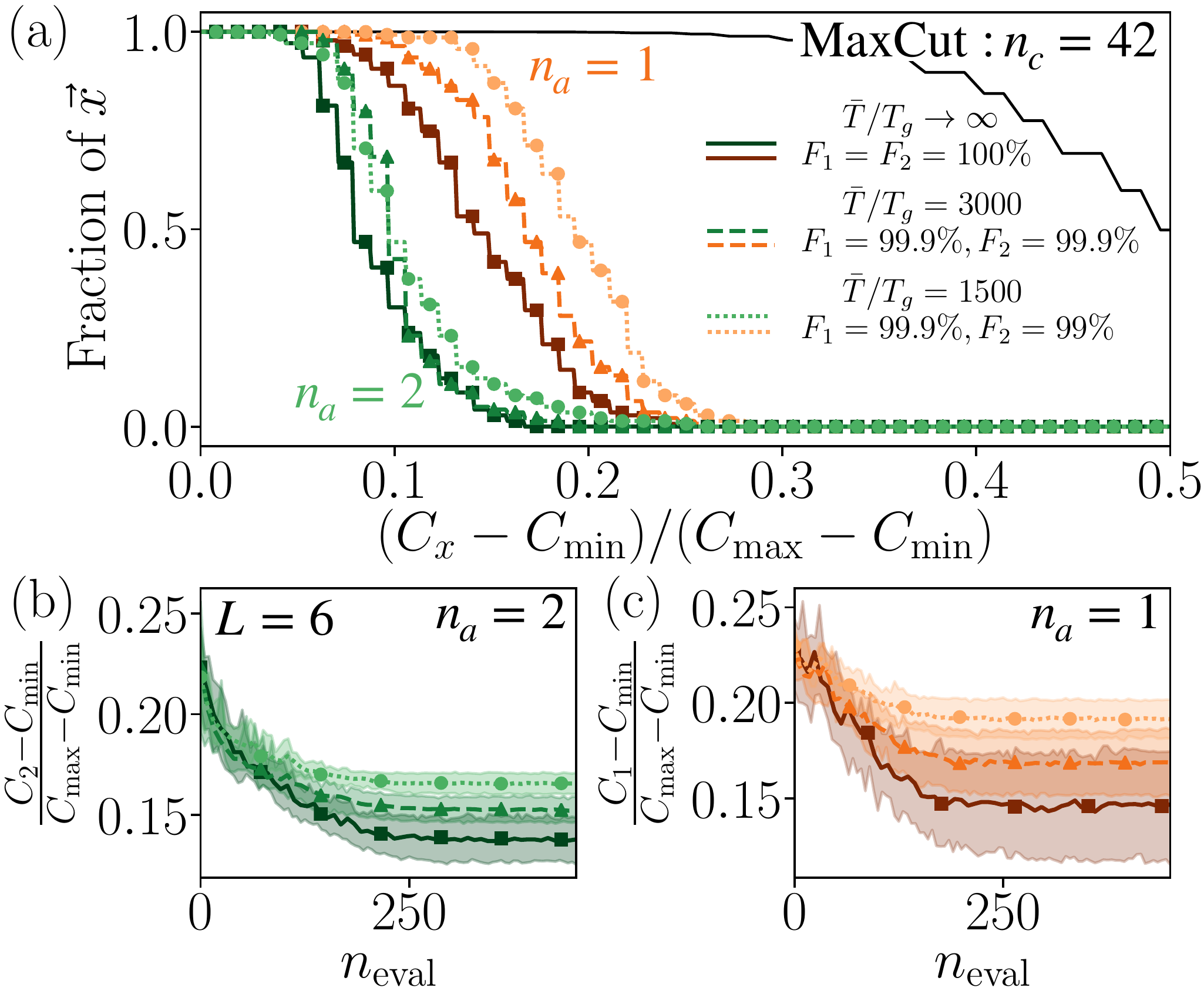}
\caption{
Solving a MaxCut problem with $n_c = 42$ in presence of noise. Reproduction of Fig.~\ref{Fig:CF_vs_L_Na2} (b) and (c) in presence of the noise models described in Appendix \ref{App:Noise} and $n_{\rm meas} = 5\times 10^4$. The protocol, QUBO matrix and optimization parameters are identical as for Fig.~\ref{Fig:CF_vs_L_Na2}.}
\label{Fig:Na2Noisy} 
\end{figure}

In Fig.~\ref{Fig:Na2Noisy}, we reproduce the results for the MaxCut problem of $n_c = 42$ variables as shown in Fig.~\ref{Fig:CF_vs_L_Na2}, this time including the noise model introduced above and using $n_{\rm meas} = 5 \times 10^4$ measurements. The comparison with the minimal encoding scheme shows an enhanced resilience to noise for the two-qubit ancilla encoding. This increased robustness could be attributed to the presence of redundancy in the encoding of correlations which can be thought of as reminiscent of the general ideas behind error encoding schemes. Such results therefore serve as additional motivation to further investigate higher-ancilla encoding schemes as they might procure additional protection against experimental imperfections.

\section{Comparison with QAOA}
\label{App:QAOA}

In this section, we compared the minimal encoding approach to the Quantum Approximate Optimization Algorithm (QAOA) under the effects of noise. QAOA is a commonly employed technique used to solve binary optimization problems on NISQ devices~\cite{Farhi2014QAOA1, Zhou_2020, Otterbach2017, Qiang2018, Pagano2019, Bengtsson2019, Abrams2019, Willsch2020, Arute2020}, where each classical variable is represented by a single qubit (complete encoding).

Using QAOA to solve a QUBO problem involves finding the state.
\begin{equation}
    \ket{\psi_{\textrm{QAOA}} (\vec{\gamma}, \vec{\beta})} = \prod_{p} \hat{U}_x(\beta_p) \hat{U}_H(\gamma_p) \ket{+}^{\otimes n_q},
\end{equation}
that minimizes Eq.~\eqref{Eq:CF_FullEnc} by finding the optimal variational parameters $\vec{\gamma}$ and $\vec{\beta}$. The unitaries $\hat{U}_H(\gamma)$ and $\hat{U}_x(\beta)$ take the form 
\begin{equation}
    \hat{U}_H(\gamma) = e^{-i \gamma \hat{H}_{\textrm{Ising}}}, \: \hat{U}_x(\beta) = e^{-i \beta \sum_{i}^{n_q} \sigma_x^{(i)}},
\end{equation}
where $\hat{H}_{\textrm{Ising}}$ is the Ising Hamiltonian described in Eq.~\eqref{Eq:HIsing}. One of the main advantages of QAOA is its guaranteed monotonic convergence to the optimal solutions as $p \rightarrow \infty$. However, the current capabilities of NISQ devices limits $p$ to small values and its performance has so far been drastically compromised when the interactions in $\hat{H}_{\textrm{Ising}}$ do not match the connectivity of the physical device. 

In what follows, we simulate the QAOA protocol in which we consider a linear topology where two-qubit operations can only be applied on qubits adjacent to each other (similar throughout the manuscript). Solving a general QUBO problem where all-to-all interactions can be encountered therefore requires a network of SWAP gates for all two-qubit $\hat{\sigma}_z^{(i)} \otimes \hat{\sigma}_z^{(j)}$ interactions in $\hat{H}_{\textrm{Ising}}$ to be implemented. 
One of the most successful experimental implementation of the QAOA protocol to date relied on an efficient decomposition of the $e^{- i \gamma \hat{\sigma}_z^{(i)} \otimes \hat{\sigma}_z^{(j)}}\cdot \textrm{SWAP}$ operations into native gates~\cite{Arute2020} and we simulate the same decomposition with the same gate fidelities and gate times. 
We note that different platforms require different gate decompositions due to the different native gate sets and efforts have been devoted to reduce the number of gates required~\cite{Lacroix_2020}.

To find the best parameters $\vec{\gamma}$ and $\vec{\beta}$, we adopt a commonly used optimization strategy that consists of (1) scanning the two-dimensional parameter space spanned by ($\gamma_1$,$\beta_1$) for $p=1$, (2) fixing ($\gamma_1$,$\beta_1$) to their optimal values (3) adding one additional layer $p \rightarrow p+1$ and repeating steps (1)-(3) until reaching the desired final depth. 
We note that techniques to reduce the required size of the search grid for the parameters associated with $p > 1$ have been proposed~\cite{Zhou_2020}.
During our simulation, the parameter scan was done over a $50\times50$ points grid ($\beta \in [0,\pi[$ and $\gamma \in [0,2\pi[$) with 50 000 measurements per point, well within the capabilities of existing hardware~\cite{Arute2020}. It is also noteworthy that for general instances of QUBO problems, $\gamma_1$ might not be bounded to the domain above. To ensure that the finite grid resolution was not a limiting factor, the optimal parameters found on the initial $50\times50$ points grid were further improved by performing an additional refined local search.

\begin{figure}
\center
\includegraphics[width=0.9\columnwidth]{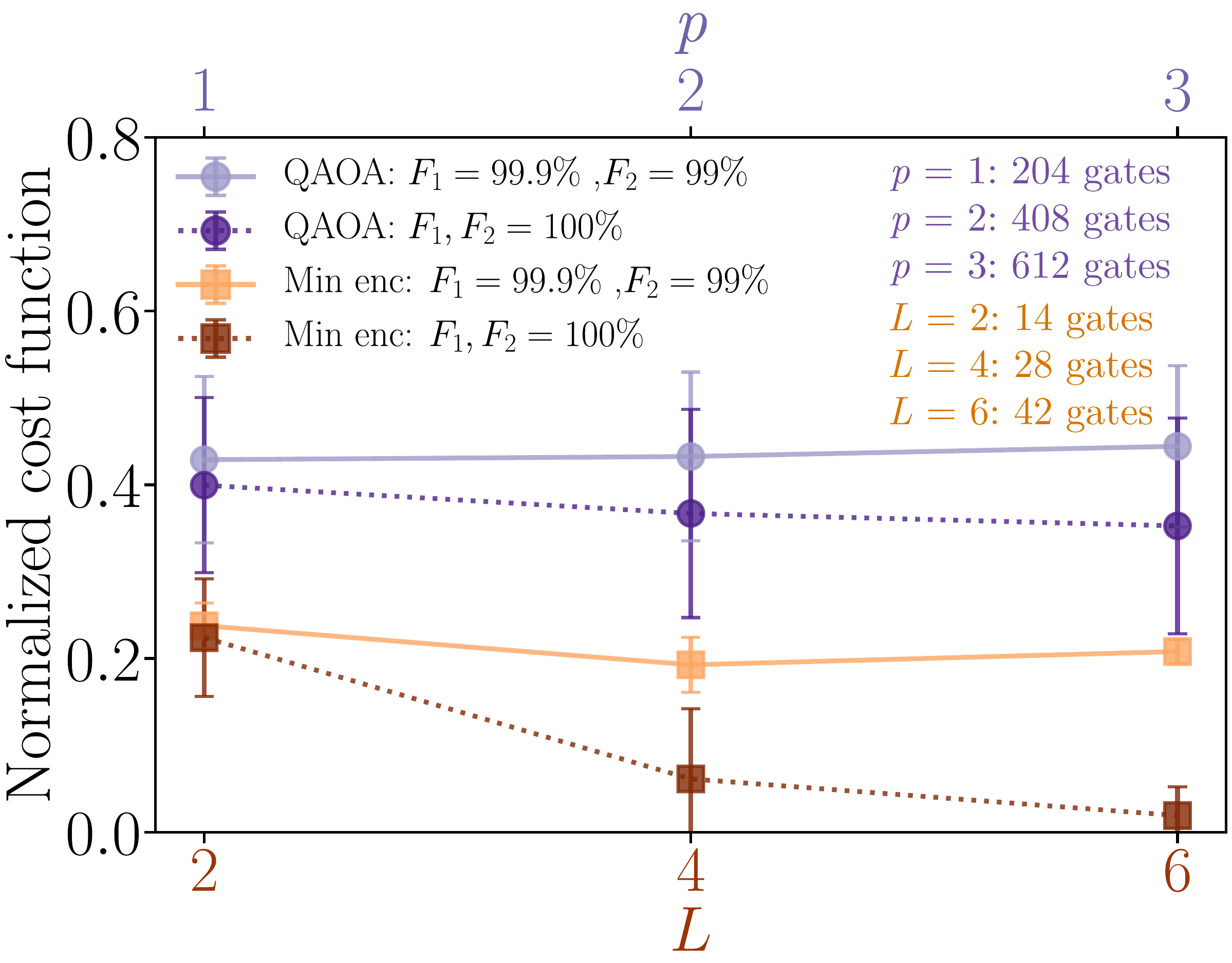}
\caption{
Noisy and noise-free simulations of QAOA and the minimal encoding scheme with hardware efficient ansatz applied to randomly generated $A$ matrices with $n_c = 8$ classical variables. Data points and error bars show the mean and standard deviation of the normalized cost function over multiple matrices (and initial parameters $\Vec{\theta}_{\textrm{ini}}$) after optimization for $p = 1, 2$ and $3$ ($L = 2, 4$ and $6$) for QAOA (minimal encoding). $8$ qubits were used in QAOA while only $4$ qubits were required in the minimal encoding scheme. For QAOA, the total fidelity of the $e^{- i \gamma \hat{\sigma}_z^{(i)} \otimes \hat{\sigma}_z^{(j)}}\cdot \textrm{SWAP}$ operations is $F=96.3\%$ with gate time $\bar{T}/T_g \approx 640$ as reported in Ref.~\cite{Arute2020}. All other parameters are identical to Fig.~\ref{Fig:CFwithnoise}.
}
\label{Fig:QAOAcomparison} 
\end{figure}

In Fig.~\ref{Fig:QAOAcomparison}, we show the comparison in the performance of our minimal encoding scheme and the QAOA protocol for multiple randomly generated $\mathcal A$ matrices of size $n_c = 8$. This is in contrast to problems artificially curated to match the topology of the quantum device commonly used in experimental implementations~\cite{Arute2020, Lacroix_2020, Willsch2020, Otterbach2017, Pagano2019}. Similar to the simulations shown in Appendix~\ref{App:Noise}, we use a noise model that, in addition to the finite gate fidelity, also includes thermal relaxations and readout errors.
We emphasize that the search protocol in both the QAOA and minimal encoding scheme have been performed in presence of the simulated noise. This is also in contrast to some recent experimental QAOA demonstrations where the optimization is first performed with an ideal simulation and only the optimized circuit is executed on the quantum hardware~\cite{Arute2020, Lacroix_2020, Willsch2020}.
For the minimal encoding, we used 15 starting points of randomly chosen parameters. With each optimization run resulting in $n_{\rm eval} \approx 200$, this leads to a similar amount of circuit evaluations equivalent to $p=1$ over a $50\times50$ points search grid.

We see from Fig.~\ref{Fig:QAOAcomparison} that despite the provable monotonic converge of the QAOA for increasing $p$, practical limitations drastically limit its application to (small) generic QUBO problems. It is therefore not surprising that our minimal encoding considerably outperforms the QAOA given the important difference in the required resources. 
For a single layer of $p=1$ for $n_c =8$, implementing $\hat U_H$ via a SWAP network over 8 qubits requires 28 $\hat{\sigma}_z^{(i)} \otimes \hat{\sigma}_z^{(j)}$-SWAP interactions arranged in 8 subsequent layers. 
Following the gate decomposition used in Ref.~\cite{Arute2020}, our implementation of QAOA required 84 2-qubit gates and 112 single qubit gates for each application of $\hat U_H$.
In contrast, using a hardware efficient ansatz of the form shown in Fig.~\ref{Fig:Circuit}, our minimal encoding requires 4 qubits, 3 CNOT gates arranged in 2 subsequent layers and 4 $R_y$ parametrized rotations for $L=1$.  

\bibliographystyle{plainnat}
\bibliography{QUBO_bibliography}

\end{document}